\documentclass{article}

\usepackage{amsmath}
\usepackage{amssymb}
\usepackage{graphicx}
\usepackage{multirow}
\usepackage{subfigure,url}
\usepackage{cite}
\usepackage[top=2in, bottom=1.5in, left=1.8in, right=1.8in]{geometry}
\usepackage{setspace}
\usepackage{lscape}

\title{The GTZAN dataset: Its contents, its faults, \\
their effects on evaluation, and its future use}

\author{Bob L. Sturm% <-this % stops a space
\thanks{B. L. Sturm is with the Audio Analysis Lab, 
AD:MT, Aalborg University Copenhagen, 
A.C. Meyers V\ae nge 15, DK-2450 Copenhagen, Denmark, 
(+45) 99407633, fax: (+45) 44651800, e-mail: bst@create.aau.dk.
He is supported in part by Independent Postdoc Grant
11-105218 from Det Frie Forskningsr\aa d;
the Danish Council for Strategic Research of the 
Danish Agency for Science Technology and Innovation 
in project CoSound, case no. 11-115328.}}%

\begin{document}

%{\cmjTitle The GTZAN dataset: Its contents, faults, 
%and their effects on music genre recognition evaluation}
%\vspace*{24pt}

\maketitle

\begin{abstract}
The {\em GTZAN} dataset
appears in at least 100 published works,
and is the most-used public dataset
for evaluation in machine listening research for music genre recognition (MGR).
Our recent work, however, shows {\em GTZAN} 
has several faults (repetitions, mislabelings, and distortions), 
which challenge the interpretability
of any result derived using it.
In this article, we disprove the claims that
all MGR systems are affected in the same ways by these faults, 
and that the performances of MGR systems in GTZAN 
are still meaningfully comparable since they all face the same faults.
We identify and analyze the contents of {\em GTZAN},
and provide a catalog of its faults.
We review how {\em GTZAN} has been used in MGR research,
and find few indications that its faults have been known and considered.
Finally, we rigorously study the effects of its faults on evaluating five different MGR systems.
The lesson is not to banish {\em GTZAN},
but to use it with consideration of its contents.
\end{abstract}

%\begin{IEEEkeywords}
%Evaluation, machine listening, music genre recognition, data
%%EDICS: AUD-CONT Content-Based Music Processing;
%%AUD-ANSY Audio Analysis and Synthesis
%\end{IEEEkeywords}

%\vspace{-0.1in}
\section{Introduction}
Our recent review of over 467 published works in music genre recognition (MGR) \cite{Sturm2012d}
shows that the most-used public dataset
is {\em GTZAN},\footnote{\url{http://marsyas.info/download/data_sets}}
appearing in the evaluations of 100 works %since its creation in 2002 
\cite{Ahonen2010,Anden2011,Anglade2010,Arabi2009,Ariyaratne2012,Bagci2007,Barreira2011,Behun2012,Benetos2008,Benetos2010,Bergstra2006,Bergstra2006b,Bergstra2010,Bogdanov2011,Chang2010,Chathuranga2012,Chen2006,Chen2008,Dixon2010,Draman2011,Fu2010,Fu2010b,Fu2011b,Genussov2010,Guaus2009,Hamel2010,Hartmann2011,Henaff2011,Holzapfel2007,Holzapfel2008,Karkavitsas2011,Karkavitsas2012,Kotropoulos2010,Krasser2012,Lampropoulos2010,Lee2009b,Leon2012,Leon2012b,Li2003,Li2003b,Li2005,Li2005b,Li2006,Li2010,Li2011,Lidy2005,Lidy2006,Lidy2007,Lidy2010b,Lim2011,Liu2009,Manzagol2008,Markov2012,Markov2012b,Marques2011c,Mayer2010c,Moerchen2006,Nagathil2011,Panagakis2008,Panagakis2009,Panagakis2009b,Panagakis2010,Panagakis2010c,Ravelli2010,Ren2011,Ren2012,Rocha2011,Rump2010,Salamon2012,Santos2010,Schindler2012b,Seo2011,Serra2011,Seyerlehner2010,Seyerlehner2010b,Seyerlehner2012,Shen2005,Shen2006,Sotiropoulos2008,Srinivasan2004,Sturm2012,Sturm2012c,Sturm2012e,Sturm2013,Sturm2013b,Tietche2012,Tsunoo2009,Tsunoo2009b,Tsunoo2011,Turnbull2005,Tzanetakis2002,Tzanetakis2002b,Wu2011,Wulfing2012,Yang2011b,Yaslan2006b,Yeh2012,Yeh2013,Zeng2009,Zhou2012}.
{\em GTZAN} %, created in 2002 \cite{Tzanetakis2002,Tzanetakis2002b},
is composed of 1,000 half-minute music audio excerpts
singly labeled in ten categories \cite{Tzanetakis2002,Tzanetakis2002b};
and though its use is so widespread, 
it has always been missing metadata identifying its contents.
In fact, {\em GTZAN} was not expressly created for MGR,\footnote{Personal communication with G. Tzanetakis} 
but its availability has made it 
a benchmark dataset, and thus a measuring stick
for comparing MGR systems, e.g., \cite{Fu2011}.
However, few researchers have ever listened to 
and critically evaluated the contents of {\em GTZAN},
and thus its faults remained undiscovered since its creation in 2002.
%Despite this, only recently has the
%composition of {\em GTZAN} begun to be formally and ,
%and its faults in integrity laid bare.

Our previous work \cite{Sturm2012b} 
provides the first metadata for {\em GTZAN},
and identifies several faults in its integrity:
repetitions, % of four different kinds,
mislabelings, % of two different kinds, 
and distortions.
In that paper, however, we do not find the extent to which
{\em GTZAN} appears in the literature,
and do not survey the ways in which it has been used.
We do not measure how its faults affect evaluation in MGR;
and, furthermore, we do not provide any recommendations for its future use.
This article rigorously addresses all of these,
significantly extending our analysis of {\em GTZAN}
in several practical ways.
%and provides an archival resource that surveys and
%links together a significant amount of work.
%published in this and other IEEE Transactions,
%Letters, Magazines, and affiliated conferences.
Our work not only illuminates results 
reported by a significant amount of work,
%published in this and other IEEE Transactions,
%Letters, Magazines, and affiliated conferences,
but also provides a critical piece to address
the non-trivial problems associated with evaluating
music machine listening systems in valid ways
\cite{Urbano2011b,Sturm2012b,Sturm2012c,Sturm2012d,Sturm2012e,Sturm2013e}.

As a brief menu of our main results,
the most significant one is that we disprove the claims: 
``all MGR systems are affected in the same ways by the faults in {\em GTZAN}'', 
and ``the performances of all MGR systems in {\em GTZAN}, 
working with the same data and faults, are still meaningfully comparable.''
This shows that, for the 100 works performing evaluation in {\em GTZAN}\cite{Ahonen2010,Anden2011,Anglade2010,Arabi2009,Ariyaratne2012,Bagci2007,Barreira2011,Behun2012,Benetos2008,Benetos2010,Bergstra2006,Bergstra2006b,Bergstra2010,Bogdanov2011,Chang2010,Chathuranga2012,Chen2006,Chen2008,Dixon2010,Draman2011,Fu2010,Fu2010b,Fu2011b,Genussov2010,Guaus2009,Hamel2010,Hartmann2011,Henaff2011,Holzapfel2007,Holzapfel2008,Karkavitsas2011,Karkavitsas2012,Kotropoulos2010,Krasser2012,Lampropoulos2010,Lee2009b,Leon2012,Leon2012b,Li2003,Li2003b,Li2005,Li2005b,Li2006,Li2010,Li2011,Lidy2005,Lidy2006,Lidy2007,Lidy2010b,Lim2011,Liu2009,Manzagol2008,Markov2012,Markov2012b,Marques2011c,Mayer2010c,Moerchen2006,Nagathil2011,Panagakis2008,Panagakis2009,Panagakis2009b,Panagakis2010,Panagakis2010c,Ravelli2010,Ren2011,Ren2012,Rocha2011,Rump2010,Salamon2012,Santos2010,Schindler2012b,Seo2011,Serra2011,Seyerlehner2010,Seyerlehner2010b,Seyerlehner2012,Shen2005,Shen2006,Sotiropoulos2008,Srinivasan2004,Sturm2012,Sturm2012c,Sturm2012e,Sturm2013,Sturm2013b,Tietche2012,Tsunoo2009,Tsunoo2009b,Tsunoo2011,Turnbull2005,Tzanetakis2002,Tzanetakis2002b,Wu2011,Wulfing2012,Yang2011b,Yaslan2006b,Yeh2012,Zeng2009,Zhou2012,Yeh2013}, 
one cannot make any meaningful conclusion about which
system is better than another for reproducing the labels of {\em GTZAN},
let alone which is even addressing the principal goals of MGR \cite{Sturm2013e}.
%From our analysis of its contents, including producing metadata
%that has never before been available,
%we estimate that a ``perfect'' MGR system
%will achieve in {\em GTZAN} a classification accuracy of 0.94.
We find that of these 100 works,
more appear in 2010 -- 2012
than in the eight years after its creation.
We find only five works (outside our own \cite{Sturm2012c,Sturm2012e,Sturm2013,Sturm2013b},
and \cite{Yeh2013} which references \cite{Sturm2012b})
that indicate someone has listened to some of {\em GTZAN}.
Of these, one work explicitly endorses its integrity for MGR evaluation,
while four allude to some problems.
We find no work (outside our own \cite{Sturm2012c,Sturm2012e,Sturm2013,Sturm2013b}) 
that explicitly considers {\em the musical content} of {\em GTZAN} in evaluation. %, e.g.,

The lesson of this article is not that {\em GTZAN}
should be banished, but that {\em it must be used
with consideration of its contents}.
Its faults, which are representative of data in the real-world,
can in fact be used in the service of evaluation \cite{Sturm2012e},
no matter if it is MGR, or other music machine listening tasks. %autotagging, or music similarity.

In the next subsection, we enumerate our contributions,
and then explicitly state delimitations of this article.
We then address several criticisms that have been raised
in reviews of versions of this work.
In the second section, we extend our previous analysis
of {\em GTZAN} \cite{Sturm2012b}.
In the third section, we comprehensively survey how
{\em GTZAN} has been used.
%\cite{Ahonen2010,Anden2011,Anglade2010,Arabi2009,Ariyaratne2012,Bagci2007,Barreira2011,Behun2012,Benetos2008,Benetos2010,Bergstra2006,Bergstra2006b,Bergstra2010,Bogdanov2011,Chang2010,Chathuranga2012,Chen2006,Chen2008,Dixon2010,Draman2011,Fu2010,Fu2010b,Fu2011b,Genussov2010,Guaus2009,Hamel2010,Hartmann2011,Henaff2011,Holzapfel2007,Holzapfel2008,Karkavitsas2011,Karkavitsas2012,Kotropoulos2010,Krasser2012,Lampropoulos2010,Lee2009b,Leon2012,Leon2012b,Li2003,Li2003b,Li2005,Li2005b,Li2006,Li2010,Li2011,Lidy2005,Lidy2006,Lidy2007,Lidy2010b,Lim2011,Liu2009,Manzagol2008,Markov2012,Markov2012b,Marques2011c,Mayer2010c,Moerchen2006,Nagathil2011,Panagakis2008,Panagakis2009,Panagakis2009b,Panagakis2010,Panagakis2010c,Ravelli2010,Ren2011,Ren2012,Rocha2011,Rump2010,Salamon2012,Santos2010,Schindler2012b,Seo2011,Serra2011,Seyerlehner2010,Seyerlehner2010b,Seyerlehner2012,Shen2005,Shen2006,Sotiropoulos2008,Srinivasan2004,Sturm2012,Sturm2012c,Sturm2012e,Sturm2013,Sturm2013b,Tietche2012,Tsunoo2009,Tsunoo2009b,Tsunoo2011,Turnbull2005,Tzanetakis2002,Tzanetakis2002b,Wu2011,Wulfing2012,Yang2011b,Yaslan2006b,Yeh2012,Zeng2009,Zhou2012}.
In the fourth section, we test the effects of
its faults on the evaluation of several categorically different
and state-of-the-art MGR systems.
Finally, we conclude by describing
how {\em GTZAN} can be useful to future research.
We make available on-line the MATLAB code to reproduce all results and figures, 
as well as the metadata for {\em GTZAN}:
\url{ http://imi.aau.dk/~bst}.

%\vspace{-0.1in}
\subsection{Contributions}
\begin{enumerate}

\item We evaluate for several MGR systems
the effects of the faults in {\em GTZAN}
when evaluating performance,
and falsify the claims: ``all MGR systems are affected in the same ways by the faults in {\em GTZAN}'', and ``the performances of all MGR systems in {\em GTZAN}, 
working with the same data and faults, are still meaningfully comparable.''

\item We estimate upper bounds for several figures of merit 
for the ``perfect'' MGR system evaluated using {\em GTZAN}.

\item  We significantly extend our prior analysis of {\em GTZAN} \cite{Sturm2012b}:
we create metadata for 110 more excerpts;
we devise an approach to analyze 
the composition {\em and} meaning of the categories of {\em GTZAN};
and we formally define and identify mislabelings.

\item We demonstrate how {\em GTZAN} can be useful
for future research in MGR, audio similarity, autotagging, etc.

\item We confirm the prediction of Seyerlehner \cite{Seyerlehner2010,Seyerlehner2010b} 
that {\em GTZAN} has a large amount of artist repetition;
we measure for the first time for {\em GTZAN} the effect of this 
on MGR evaluation.

\item  We provide a comprehensive survey of how 
{\em GTZAN} has been used,
which ties together 100 published works %published by IEEE
that use {\em GTZAN} for evaluation,
and thus are affected by its faults.
%seven articles in 
%TASLP \cite{Tzanetakis2002,Benetos2010,Holzapfel2008,
%Panagakis2010,Ravelli2010,Ren2012,Tsunoo2011};
%seven articles appearing in other
%Transactions, Letters, and Magazines \cite{Lee2009b,Shen2006,
%Bagci2007,Bogdanov2011,Li2006,Turnbull2005,Fu2011};
%and at least 21 papers appearing in IEEE conferences
%\cite{Arabi2009,Ariyaratne2012,Chen2006,Holzapfel2007,Li2003,Li2005,Markov2012,Markov2012b,
%Nagathil2011,Panagakis2010,Ren2011,Salamon2012,Serra2011,Srinivasan2004,
%Sturm2013,Sturm2013b,Tietche2012,Tsunoo2009,Wu2011,Zeng2009,Yeh2013}.

%\item 
\end{enumerate}

%\vspace{-0.15in}
\subsection{Delimitations}
%It is necessary to clearly state with what this article is and is not concerned.
This article is concerned only with {\em GTZAN}: its composition and faults;
its historical and contemporary use for evaluating MGR systems;
the effects of its faults on evaluating MGR systems;
and how to use it with consideration of its problems.
%\cite{Ahonen2010,Anden2011,Anglade2010,Arabi2009,Ariyaratne2012,Bagci2007,Barreira2011,Behun2012,Benetos2008,Benetos2010,Bergstra2006,Bergstra2006b,Bergstra2010,Bogdanov2011,Chang2010,Chathuranga2012,Chen2006,Chen2008,Dixon2010,Draman2011,Fu2010,Fu2010b,Fu2011b,Genussov2010,Guaus2009,Hamel2010,Hartmann2011,Henaff2011,Holzapfel2007,Holzapfel2008,Karkavitsas2011,Karkavitsas2012,Kotropoulos2010,Krasser2012,Lampropoulos2010,Lee2009b,Leon2012,Leon2012b,Li2003,Li2003b,Li2005,Li2005b,Li2006,Li2010,Li2011,Lidy2005,Lidy2006,Lidy2007,Lidy2010b,Lim2011,Liu2009,Manzagol2008,Markov2012,Markov2012b,Marques2011c,Mayer2010c,Moerchen2006,Nagathil2011,Panagakis2008,Panagakis2009,Panagakis2009b,Panagakis2010,Panagakis2010c,Ravelli2010,Ren2011,Ren2012,Rocha2011,Rump2010,Salamon2012,Santos2010,Schindler2012b,Seo2011,Serra2011,Seyerlehner2010,Seyerlehner2010b,Seyerlehner2012,Shen2005,Shen2006,Sotiropoulos2008,Srinivasan2004,Sturm2012,Sturm2012c,Sturm2012e,Sturm2013,Sturm2013b,Tietche2012,Tsunoo2009,Tsunoo2009b,Tsunoo2011,Turnbull2005,Tzanetakis2002,Tzanetakis2002b,Wu2011,Wulfing2012,Yang2011b,Yaslan2006b,Yeh2012,Zeng2009,Zhou2012}.
This article is not concerned with other public datasets,
which may or may not suffer from the same faults as {\em GTZAN},
but which have certainly been used in fewer published works than {\em GTZAN} \cite{Sturm2012d}.
This article is not concerned with the validity, well-posedness,
value, usefulness, or applicability of MGR;
or whether MGR is ``replaced by,'' or used in the service of,
e.g., music similarity, autotagging, or the like.
It is not concerned with making general conclusions about
or criticisms of MGR or evaluation in MGR, 
which are comprehensively addressed
in other works, e.g., \cite{McKay2006,Craft2007,Craft2007b,Wiggins2009,
Sturm2012c,Sturm2012d,Sturm2012e,Sturm2013e}.
Finally, it is not concerned with how,
or even whether it is possible, 
to create faultless datasets for MGR, 
music similarity, autotagging, and the like.

%\vspace{-0.15in}
\subsection{Criticisms}
A variety of criticisms have been raised in reviews of this work.
%from its usefulness, novelty and technical content,
%to the aims and use of MGR. %, the requirements of IEEE transactions,
%and so on.
%We now address the variety of these arguments.
%Though we have unsuccessfully submitted portions of this material to TASLP,
%and related conferences,
%only a portion of the current Section II
%has been accepted as a conference paper \cite{Sturm2012b}.
%Through this history then,
%s
First, its usefulness has been challenged:
``recent publications are no longer using {\em GTZAN}'';
or, ``it is already a commonly accepted opinion that the GTZAN dataset should be discarded'';
or, ``better datasets are available.''
The fact is, of all datasets that are publicly available, {\em GTZAN} is the one 
appearing most in published MGR work \cite{Sturm2012d}.
The fact is, more papers use {\em GTZAN} in the past three years
than in the first eight years of its existence (Fig. \ref{fig:GTZANstats}),
and none of them mention a ``common opinion''
that {\em GTZAN} should be discarded.
%(14 of 99 works appearing in IEEE archival formats).
Even with its faults, the fact is that {\em GTZAN} can be useful \cite{Sturm2012e,Sturm2013b}, 
and there is no reason to discard it.
Our work provides novel insights and perspectives essential 
for interpreting all published results that use {\em GTZAN},
in the past and in the future,
and ways for better scientific evaluation using it.
These facts provide strong argumentation for the 
publication and preservation of this work in an archival form.

One might challenge the aims of published work that uses {\em GTZAN}:
``to some extent genre classification has been replaced 
by the more general problem of automatic tagging'';
or, ``genre classification is now critically seen
even by the person who compiled {\em GTZAN} 
(personal communication with Tzanetakis).''
The fact is, %regardless of what one person thinks of MGR,
researchers have published and still are publishing a large number of works in MGR that use {\em GTZAN}.
One might argue, ``this article only touches the possibility of a critique of genre classification'';
or, ``it misses the valuable chance to criticize the simplistic and superficial approach 
to music that governs most of these publications'';
or, ``it falls short of providing a more rigorous questioning 
of MGR evaluation and a more general evaluation 
that goes beyond the {\em GTZAN} dataset.''
Such aspects, however,
are outside the scope of this article delimited above,
but are thoroughly addressed by
other work, e.g., \cite{Craft2007,Craft2007b,Wiggins2009,Sturm2012d,Sturm2012c,Sturm2012e,Sturm2013e}.
\section{Analysis of {\em GTZAN}}
To analyze {\em GTZAN}, we first identify its excerpts,
%(We have also listened to every excerpt multiple times.)
see how artists compose each category,
and survey the ways people %have used tags to 
describe the music. %and/or artist.
%from which each excerpt is taken, or the artist that created the music.
Finally, we delimit and identify three kinds of faults in the dataset:
repetitions, mislabelings, and distortions (summarized in Table \ref{tab:problems1}).

%\vspace{-0.15in}
\subsection{Identifying excerpts}
We use the Echo Nest Musical Fingerprinter (ENMFP)\footnote{\url{http://developer.echonest.com}} %\footnote{http://developer.echonest.com/docs/v4}
to generate a fingerprint of every excerpt in {\em GTZAN}
and to query the Echo Nest database having over 30,000,000 songs.
The second column of
Table \ref{tab:FPIDed} shows that this identifies
only 60.6\% of the excerpts.
%When it finds a match, ENMFP returns the artist and title of the excerpt, 
%which we find in many cases to be inaccurate ---
%especially for classical music, and songs from compilation albums.
We correct titles and artists as much as possible,
%e.g., we reduce ``River Rat Jimmy (Album Version)'' to ``River Rat Jimmy''; 
%and ``Bach - The \#1 Bach Album (Disc 2) - 13 - Ich steh mit einem Fuss im Grabe, BWV 156 Sinfonia''
%to ``Ich steh mit einem Fuss im Grabe, BWV 156 Sinfonia;''
%and we correct ``Leonard Bernstein [Piano], Rhapsody in Blue''
%to ``George Gershwin'' and ``Rhapsody in Blue.''
and find four misidentifications.
%Country 15\footnote{This is the file ``country.00015.wav'' in {\em GTZAN}.} 
%is misidentified as being by Waylon Jennings (it is by George Jones);
%Pop 65 is misidentified as being Mariah Carey (it is Prince); 
%Disco 79 is misidentified as
%``Love Games'' by Gazeebo (it is ``Love Is Just The Game'' by Peter Brown);
%and Metal 39 is Metallica playing ``Star Wars Imperial March,''
%but ENMFP identifies it as a track on a CD for improving sleep.\footnote{
%``Power Nap'' by J. S. Epperson (Binaural Beats Entrainment), 2010.}
We then manually identify 313
of the remaining excerpts. %, e.g.,
%by querying song lyrics on Google\footnote{A WWW search engine \url{http://google.com}}
%and confirming using YouTube,\footnote{A WWW video sharing service \url{http://www.youtube.com}} 
%or {\tt last.fm};\footnote{A crowd-sourcing WWW music business gathering 
%metadata, listening habits, etc. of a diverse on-line community \url{http://last.fm}}
%by finding track listings from online music retailers\footnote{Such as the WWW
%media retailer \url{http://amazon.com}} 
%(when it is clear the excerpts are ripped from an album),
%and confirming by listening to the available excerpts; by asking other people;
%or by using the music identification service Shazam.\footnote{A WWW
%music information business providing a mobile fingerprinting service \url{http://www.shazam.com}}
The third column of Table \ref{tab:FPIDed} shows 
that we miss the metadata of only 81 excerpts.
%A machine-readable index of the artists and titles is available at: \url{http://imi.aau.dk/~bst}.
%most of which are in the categories Jazz, Reggae, and Metal.
%Figure \ref{fig:GTZANartists} shows how multiple artists
%compose each category in {\em GTZAN}.
%We see only nine artists are represented in the Blues category.
%The category with the most artist diversity appears to be Disco,
%where we find at least 55 different artists.
%Disco is also the category with the fewest number of excerpts
%by a single artist: 7 excerpts by KC and The Sunshine Band.
%All other categories have at least 8 excerpts by one artist.
%Reggae is the category with the most excerpts from a single artist:
%35 excerpts of Bob Marley.
With this, we can now bound the number of artists in {\em GTZAN} ---
which has been stated to be unknown, e.g., \cite{Seyerlehner2010,Seyerlehner2010b}.
Assuming each of the unidentified excerpts come from different artists
than those we have identified,
the total number of artists represented in the excerpts of {\em GTZAN}
cannot be larger than 329.
If all unlabeled excerpts are from the 
same artists we have identified,
then the smallest it can be is 248.

\begin{table}[t]
%\tiny
\scriptsize
\centering
\begin{tabular}{|r|c|c|c|c|}
\hline
  &  & & \multicolumn{2}{c|}{{\tt last.fm}} \\
 {\bf Label} & {\bf ENMFP} & {\bf self} & {\bf song (no. tags)} & {\bf artist (no. tags)} \\
\hline
{\em Blues} & 63 & 100 & 75 (2904) & 25 (2061)\\
\hline
{\em Classical} & 63 & 97 & 12 (400) & 85 (4044) \\
\hline
{\em Country} & 54 & 95 & 81 (2475) & 13 (585) \\
\hline
{\em Disco} & 52 & 90 & 82 (5041) & 8 (194) \\
\hline
{\em Hip hop} & 64 & 96 & 90 (6289) & 5 (263) \\
\hline
{\em Jazz} & 65 & 80 & 54 (1127) & 26 (2102) \\
\hline
{\em Metal} & 65 & 83 & 73 (5579) & 10 (786) \\
\hline
{\em Pop }& 59 & 96 & 89 (7173) & 7 (665) \\
\hline
{\em Reggae} & 54 & 82 & 73 (4352) & 9 (616) \\
\hline
{\em Rock} & 67 & 100 & 99 (7227) & 1 (100) \\
\hline
\hline
Total & 60.6\% & 91.9\% & 72.8\% (42567) & 18.9\% (11416) \\
\hline
\end{tabular}
\caption{For each category of {\em GTZAN}:
number of excerpts we
identify by fingerprint (ENMFP);
then searching manually (self);
number of songs in {\tt last.fm} (and number of tags having ``count'' larger than 0);
for tracks not found,
number of artists in {\tt last.fm} (and number of tags having ``count'' larger than 0).
Retrieved Dec. 25, 2012, 21h.}
\label{tab:FPIDed}
%\vspace{-0.15in}
\end{table}

%\vspace{-0.15in}
\subsection{Describing excerpts and categories}
To survey the ways people describe the music or artist
of each excerpt, we %use the index we create above
query the application programming interface provided by {\tt last.fm},
and retrieve the ``tags'' users of the service
apply to the songs or artists in {\em GTZAN}.
A tag is a word or phrase a person 
uses to describe an artist, or a song, in order to make their
music collection more useful for them.
On {\tt last.fm}, tags are most often genre labels (``Blues'') \cite{Bertin-Mahieux2010c},
but they can also describe instrumentation (``female vocalists''), 
tempo (``120 bpm''), mood (``happy''), how the music is used (``exercise''),
lyrics (``fa la la la la''), reproduce the band name (``The Rolling Stones''),
or something else (``favorite song of all time'') \cite{Law2011}.
We assume the tags users give to a song or artist
would also be given to the excerpt in {\em GTZAN}.
Tags from {\tt last.fm} have been used in other work, e.g., \cite{Bertin-Mahieux2008,Barrington2008}.
%and 2) that the tags on {\tt last.fm}
%for the most part reflect the qualities
%of the underlying music or artist.

The collection of these tags is of course far from being controlled;
but an aspect that helps strengthen their use is that
{\tt last.fm} provides a ``count'' for each one: a normalized quantity such that
100 means the tag is applied by most listeners,
and 0 means the tag is applied by the fewest.
We keep only those tags with counts greater than 0.
The fourth column of Table \ref{tab:FPIDed} shows 
the number of tracks we identify that have tags in {\tt last.fm},
and the number of tags with non-zero count.
When we do not find tags for a song,
we get the tags applied to the artist.
For instance, though we identify all 100 excerpts in Blues,
only 75 of the songs are tagged. % in {\tt last.fm}.
Of these, we get 2,904 tags with non-zero counts.
For the remaining 25 songs,
we retrieve 2,061 tags from the tags given to the artists.
%and not to determine any ``true'' genre of

\newgeometry{top=1in, bottom=1in, left=0.5in, right=0.5in}
\begin{figure*}[t]
\centering
\subfigure{
\includegraphics[width=1\columnwidth]{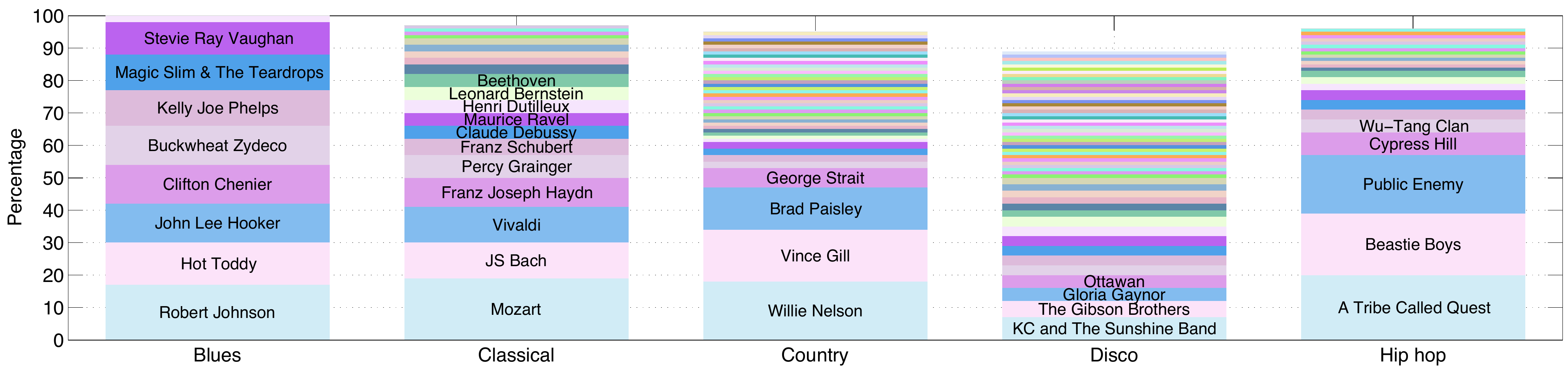}}
\subfigure{
\includegraphics[width=1\columnwidth]{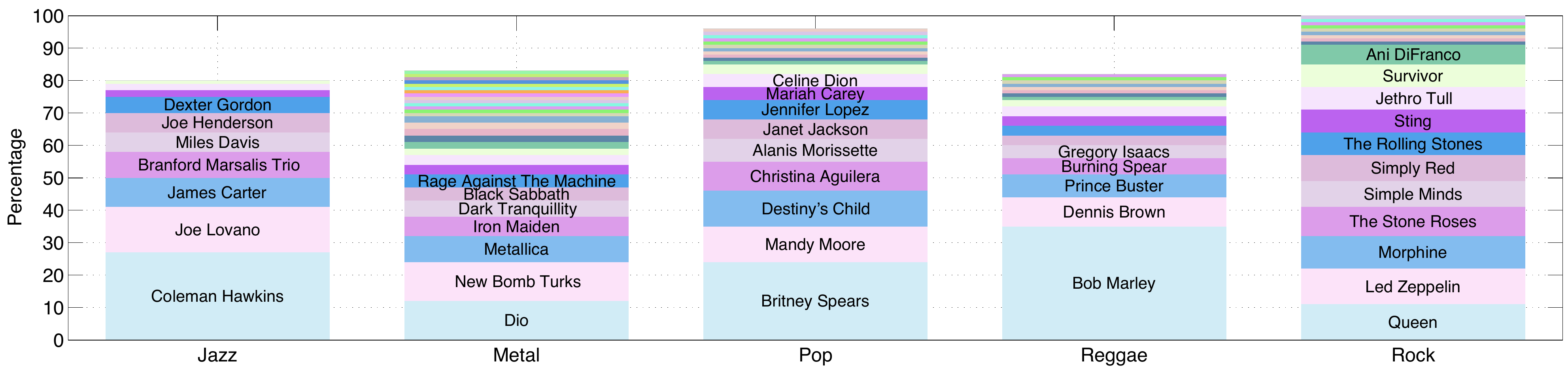}}
%\vspace{-0.1in}
\caption{Artist composition of each {\em GTZAN} category.
We do not include unidentified excerpts. %, e.g., in Jazz there remain
%20 unidentified artists.
%We only label those artists having more than 3 excerpts in each category.
}
\label{fig:GTZANartists}
%\vspace{-0.06in}
\end{figure*}

\begin{figure*}[t]
\centering
\subfigure{
\includegraphics[width=1\columnwidth]{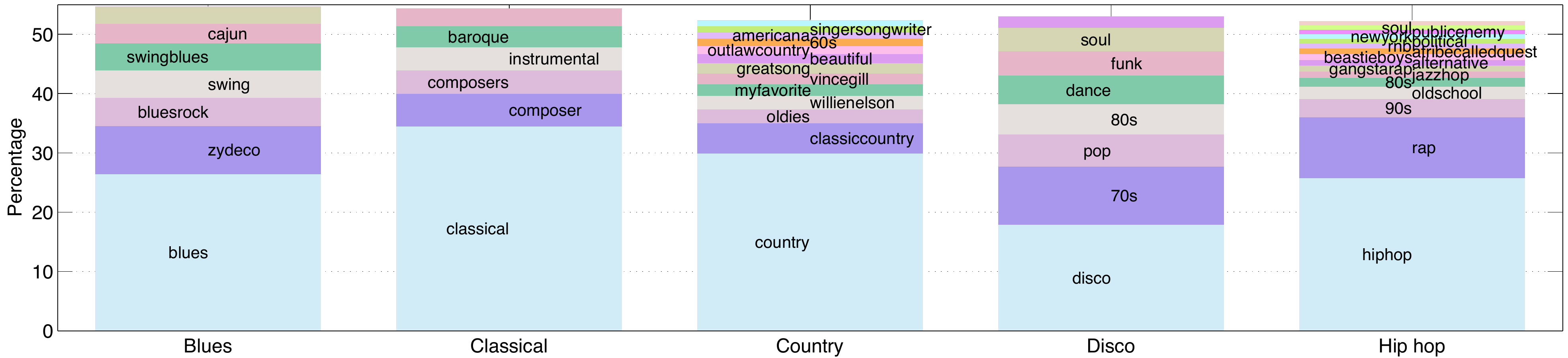}}
\subfigure{
\includegraphics[width=1\columnwidth]{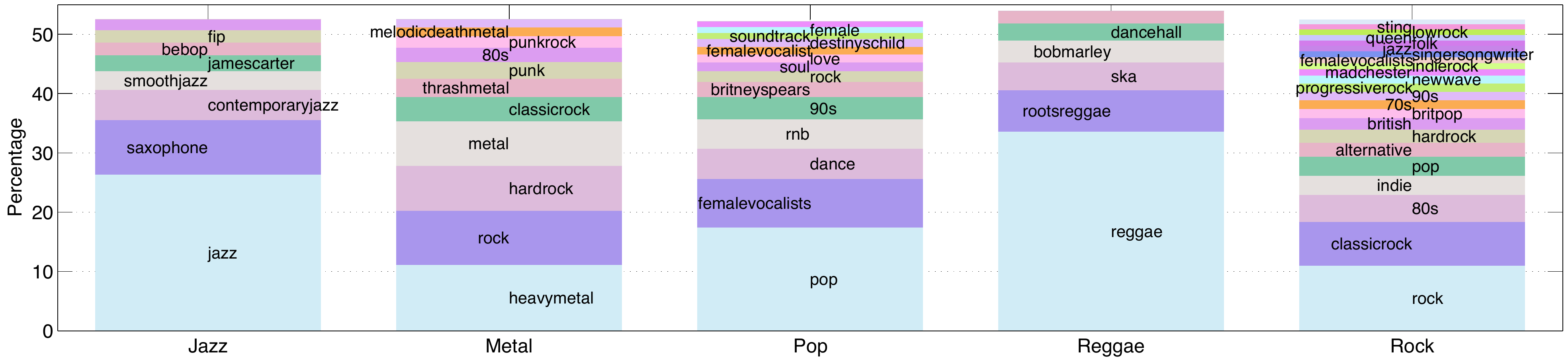}}
%\vspace{-0.1in}
\caption{Top tags of each {\em GTZAN} category.
We do not include unidentified excerpts.}
\label{fig:GTZANtags}
%\vspace{-0.15in}
\end{figure*}
\restoregeometry

%they are at the moment the best we have for large collections of music.
%Furthermore, it has been noted that a majority (68\%) of the tags on 
%{\tt last.fm} are genre \cite{Bertin-Mahieux2010c}.

%and remove all spaces from tags when they consist of multiple words,
%e.g., ``My favorite song'' becomes ``myfavoritesong.''

%For 919 identified excerpts in {\em GTZAN}, 
%we have a total of 54,080 tags with non-zero count.
%Of those artists we have identified, 
%{\tt last.fm} is missing tags for only two:
%Nina Martinique (Country 14),
%and Dirt (Hip hop 28).

%We review the tags for mistaken identity.
%and find that 12 of 13 excerpts by the Canadian folk group
%Hot Toddy (Blues 85-97) 
%are confused with a ``deep house'' electronic dance music group named the same,
%and tagged as such. 
%The Hot Toddy song ``T-bone Shuffle'' (Blues 97), however,
%is correctly tagged,
%and so we apply those tags to the 12 misidentified excerpts.
%Finally, Hip hop 28 is ``Divine Lines''
%by the Christian rap group Dirt;
%however, since {\tt last.fm} has no tags for this track, 
%our program applies the tags for the punk group named Dirt.
%We thus ignore the tags for this excerpt.
%In some cases, the tags applied to a song do not make sense,
%and so we review every instance before confirming it to be mislabeled.
%For instance, Blues 1 is 
%``I'm in the Mood for Love'' performed by John Lee Hooker,
%but it is tagged only ``bonnie raitt'' with a count 100.
%We consider such problems when we look at mislabelings below.

%We order the tags for an excerpt based on count,
We devise the following novel approach to identify
the {\em top tags} of a music excerpt.
%First, we compute the {\em normalized counts} for the set of tags
%applied to the excerpt.
Let \(\omega_i\) be its \(i\)th tag, and \(c_i\) be its count.
Given the set of \(|\mathcal{I}|\) tag-count pairs 
\(\{(w_i,c_i)\}_{i\in \mathcal{I}}\), %where \(100 \ge c_i \ge c_{i+1} > 0\)
%for all \(i\in \mathcal{I}\),
%the normalized count of the \(j\)th tag is defined
%\begin{equation}
%\bar c(w_j) := c_j \Bigr/ \sum_{i\in \mathcal{I}} c_i.
%\end{equation}
we call the {\em top tags} of an excerpt \(\mathcal{T} := \{(w_j,c_j)\}_{j \in \mathcal{J} \subseteq \mathcal{I}}\)
those that contribute the majority of the total tag count, i.e., %\vspace{-0.09in}
\begin{multline}
\mathcal{J} := \arg \min_{\mathcal{J}' \subseteq \mathcal{I}} |\mathcal{J}'| \; \textrm{subject to} 
\min_{j\in \mathcal{J}'}\{ c_j\} > \max_{i \in \mathcal{I}\backslash\mathcal{J}'} \{ c_i\},
\sum_{j \in \mathcal{J}'}c_j > \frac{1}{2}\sum_{i\in \mathcal{I}} c_i.
\label{eq:toptags}
\end{multline}
For example, consider an excerpt has the set of tag-count pairs:
\(\{(\textrm{``folk''}, 11),\)
\((\textrm{``blues''}, 100), (\textrm{``blues guitar''}, 90)\}\).
%We call the {\em normalized count} of ``blues'' as $100/(100+90+11) = 0.497$.
Its total tag count is 201, and so its top tags are ``blues'' and ``blues guitar'',
which account for \(94.6\% = 100(100+90)/201$ of the total tag count.
We argue that the top tags of an excerpt 
give a satisfactory description of music content
since they are given by the majority of users 
who have tagged the piece of music.

In our previous analysis of {\em GTZAN} \cite{Sturm2012b},
we assume that since there is a category named, e.g., ``Country,''
then the content of those excerpts should 
possess typical and distinguishing characteristics of music using the country genre \cite{Ammer2004}:
stringed instruments such as guitar, mandolin, banjo; 
emphasized ``twang'' in playing and singing;
lyrics about patriotism, hard work and hard times; and so on.
This leads us to claim in \cite{Sturm2012b} that at least seven excerpts are mislabeled ``Country''
because they have few of these characteristics.
In fact, we find other work that %using {\em GTZAN} and other datasets
assumes the concepts of two music genre datasets overlap because
they share the same labels, e.g., the taxonomies in \cite{Moerchen2006,Guaus2009}.
We make no such assumption here, 
and instead consider a label of {\em GTZAN} to represent
the set of top tags of the set of excerpts labeled so.
%for the set of stylistic descriptors are
%used by the content of the excerpts in the category.

%By considering the tags of all identified excerpts of a category,
%we define in the same way the top tags of that category.
%and uncover these descriptors by looking at the top tags 
%of the excerpts in the category as we describe above.
To find the top tags of the set of excerpts with a label,
we first construct for the set of tag-count pairs $\{(w_k, c_k))\}_{ k \in \mathcal{K}}$
by summing the counts for all \(|\mathcal{K}|\) unique tags 
of the identified excerpts.
We then compute the normalized count of each tag:
\begin{equation}
\bar c(w_k) := c_k \Bigr/ \sum_{k\in \mathcal{K}} c_k
\end{equation}
and finally define the top tags of this category by (\ref{eq:toptags}).
%i.e., those with the greatest normalized counts that sum to over 0.5.
Figure \ref{fig:GTZANtags} shows the top tags for each category,
where we can see that most of the top tags are indicative of genre.
%The sanity of our definition of top tags is thus supported since
%most tags in {\tt last.fm} are genre \cite{Bertin-Mahieux2010c},
%and people often use genre to describe music \cite{Law2011}.
Now, we can see how {\em GTZAN} Blues 
means more than the blues of, e.g., Robert Johnson and John Lee Hooker;
and how 
%
%the top tags in Blues are: \(\{\)``blues'',
%``zydeco'', ``swing'', ``swingblues'', ``bluesrock'', ``cajun''\(\}\).
%We find in Blues 24 excerpts of zydeco music by Clifton Chenier and Buckwheat Zydeco.
%Zydeco can combine elements of blues, e.g., the twelve bar structure, 
%with elements of cajun music, e.g., use of the accordion \cite{Ammer2004}.
 {\em GTZAN} Disco is more broad than
the dance music from the seventies \cite{Ammer2004,Shapiro2005}.
%We can also see that between most categories
%there is little overlap in tags.
%with the exception of Metal and Rock 
%(which share the top tags ``rock,'' ``classic rock,'' 
%and ``hard rock''),
%and Disco and Pop (which share the top tags
%``pop,'' ``dance'' and ``rnb'').

\subsection{Identifying faults: Repetitions}
We consider four types of repetition,
from high to low specificity:
exact, recording, artist, and version.
We define an {\em exact repetition} as when 
two excerpts are the same to such a degree that
their time-frequency fingerprints are highly similar.
To find exact repetitions,
we implement a simplified version of the Shazam fingerprint \cite{Wang2003},
compare every pair of excerpts,
and then listen to confirm.
The second column of Table \ref{tab:problems1} 
shows these %Jazz category has the most exact repetitions (13),
%followed by Reggae (11) and Pop (9).
%We find no exact repetitions in the 
%Blues, Classical and Country.
In total, we find 50 exact repetitions in {\em GTZAN}.

We define a {\em recording repetition} as when 
two excerpts come from the same recording, 
but are not detected with fingerprinting.
%perhaps only displaced in time.
We find these by looking for artists and songs appearing multiple times
in the metadata.
To confirm, we listen to the excerpts. %and identify from where they come
%in the original recording.
%For instance, Country 8 and 51 
%are both from ``Never Knew Lonely'' by Vince Gill,
%but excerpt 8 comes from later in the recording than excerpt 51.
The second and third columns of Table \ref{tab:problems1} shows these. %the category
We find Pop has the most exact and recording repetitions (16):
``Lady Marmalade'' sung by Christina Aguilera et al.,
as well as ``Bootylicious'' by Destiny's Child,
each appear four times.
%We also find Reggae 85 is ``Judge Dread''
%and Reggae 96 is ``Barrister Pardon,'' both by Prince Buster,
%are similar enough to be considered to come from the same recording.
%(The tempo of the latter is higher, but the instrumentation,
%rhythm, and feel are the same.)
In total, we find 21 recording repetitions in {\em GTZAN}.

We define {\em artist repetition} as excerpts 
performed by the same artist,
which are easily found using the metadata.
%Figure \ref{fig:GTZANartists} and 
Table \ref{tab:problems1} shows how 
every category of {\em GTZAN} has artist repetition.
We see the 100 excerpts in Blues come from only nine artists;
and more than a third of the excerpts labeled Reggae 
come from Bob Marley.
%The category with the least number of artists is Blues (9),
%and with the most number of identified artists is Disco (55).
%Three excerpts in Metal, and 11 in Rock,
%come from the group Queen.
%Two excerpts in Metal and one in Rock come from Guns 'N Roses. 

%\begin{figure}[t]
%\centering
%\includegraphics[width=0.4\columnwidth]{figures/fingerprints/fingerprints}
%\vspace{-0.1in}
%\caption{Exact repetitions appear clearly 
%in this portion of the pair-wise hash comparisons in
%the {\em GTZAN} excerpts labeled Jazz.}
%\label{fig:fingerprints}
%\vspace{-0.2in}
%\end{figure}

We define a {\em version repetition} as when 
two excerpts are of the same song but performed differently.
This could be a studio version, a live version,
performed by the same or different artists,
or even a remix.
We identify these with the metadata,
and then confirming by listening.
For instance, Classical 44 and 48 are
from ``Rhapsody in Blue'' by George Gershwin,
but performed by different orchestras.
%Disco 66 and 69 are both
%``Shake Your Groove Thing'' performed by
%Peaches and Herb, but in different ways.
%Excerpts 2 and 32 in Hip hop are both
%versions of ``Buddy'' by De La Soul.
Metal 33 is ``Enter Sandman'' by Metallica, 
and Metal 74 is a parody. %``Internet Sandman'' by Johnny Crass.
%And Metal 85 is Ozzy Osbourne covering
%``Staying Alive'' by the Bee Gees, 
%an excerpt of which appears as Disco 14.
%The category with the most number of 
%version repetitions is Pop (6).
%Pop 16 and 17 are both
%``I can't get no satisfaction'' by Britney Spears,
%but the latter is performed live.
%Pop 75 and 82 are ``If you had my love'' by Jennifer Lopez,
%but the latter is an uptempo Latin remix.
%And Reggae 23 is ``Sun is Shining'' by Bob Marley, 
%while Reggae 55 is a dance remix of the same.
In total, we find 13 version repetitions in {\em GTZAN}.

\subsection{Identifying faults: Mislabelings}\label{sec:mislabelings}
In \cite{Sturm2012b}, we consider two kinds of mislabelings: 
contentious and conspicuous. 
We based these upon non-concrete criteria
formed loosely around musicological principles 
associated with the genre labels of {\em GTZAN} ---
which we have shown above are not indicative of the content of each category.
Here, we formalize the identification of mislabeled excerpts
by using the concept of top tags we develop above.
%Essentially, we say an excerpt is {\em mislabeled}
%if we find little overlap between
%the tags of the excerpt and the top tags of its label in {\em GTZAN}.
%To measure this overlap, we compare how
%tags are shared between the excerpt and each category.

Consider an excerpt with label $g \in \mathcal{G}$ has the set of tag-normalized count pairs
$\mathcal{X} = \{(w_i,\bar c_i(w_i))\}_{i \in \mathcal{I}}$.
We know label $r \in \mathcal{G}$ has the set of 
top tag-normalized count pairs
$\mathcal{T}_r = \{(y_j, \bar d(y_j))\}_{ j \in \mathcal{J}_r}$.
We define the {\em \(r\)-label score of $\mathcal{X}$}
\begin{equation}
C(\mathcal{X}, \mathcal{T}_r) := \sum_{j \in \mathcal{J}_r} \bar d_j(y_j) \sum_{i \in \mathcal{I}}  \bar c_i(\omega_i) \delta_{y_j \equiv w_i}
\label{eq:score}
\end{equation}
where $\delta_{y_j \equiv w_i} = 1$ if
$y_j$ and $w_i$ are identical,
or zero otherwise.
Note, this compares all tags of an excerpt
to only the top tags of label $r$.
Now, if $C(\mathcal{X},\mathcal{T}_g)$ is {\em too small}
(the tags of the excerpt have too little in common 
with the top tags of its label),
or if $C(\mathcal{X},\mathcal{T}_r)$ is {\em too large} for an \(r \ne g\)
(the tags of the excerpt have more in common with the top tags of another label),
then we say the excerpt is mislabeled.

\newgeometry{top=1in, bottom=1in, left=0.3in, right=0.1in}
\begin{table*}[htbp]
\singlespacing
%\centering
%\tiny
\scriptsize
\begin{tabular}{|@{\;}c@{\;}|@{\,}p{0.5in}@{\;}|@{\,}p{0.5in}@{\;}|@{\,}p{2.5in}@{\;}|@{\,}p{0.4in}@{\;}|@{\,}p{2.5in}@{\;}|@{\;}p{0.6in}@{\;}|}
\hline
{\bf Label}  & \multicolumn{4}{c|@{\,}}{{\bf Repetitions}} & \centering {\bf Mislabelings} & {\bf Distortions} \\
 & \centering{\em Exact} & \centering {\em Recording} & \centering {\em Artist} & {\em Version} &   & \\
\hline
{\em Blues} &
      &  &  John Lee Hooker (0-11); Robert Johnson (12-28); Kelly Joe Phelps (29-39); 
     Stevie Ray Vaughn (40-49); Magic Slim (50-60); Clifton Chenier (61-72);
     Buckwheat Zydeco (73-84); Hot Toddy (85-97); Albert Collins (98, 99) &  &    
     & \\
\hline
{\em Classical} &
      & (42,53) (51,80) & 
      J. S. Bach (00-10); Mozart (11-29); Debussy (30-33); Ravel (34-37); Dutilleux (38-41); Schubert (63-67);
      Haydn (68-76); Grainger (82-88); Vivaldi (89-99); and others &  (44,48) &   & static (49) \\
\hline
{\em Country} &
      &  (08,51) (52,60) & Willie Nelson (19,26,65-80); Vince Gill (50-64); Brad Paisley (81-93); 
      George Strait (94-99); and others & (46,47)  & 
      %Burt Bacharach ``Raindrops Keep Falling on my Head'' (21);
      Wayne Toups \& Zydecajun ``Johnnie Can't Dance'' (39)      
       & static distortion (2) \\
\hline
{\em Disco} &
      (50,51,70) (55,60,89) (71,74) (98,99) &  (38,78) & Gloria Gaynor (1,21, 38,78); 
      Ottawan (17,24,44,45); The Gibson Brothers (22,28,30,35,37); 
      KC and The Sunshine Band (49-51,70,71,73,74); ABBA (67,68,72); and others &  (66,69) &
      Billy Ocean ``Can You Feel It'' (11); 
      Clarence Carter ``Patches" (20);
      Latoya Jackson ``Playboy" (23), ``(Baby) Do The Salsa" (26); 
      The Sugarhill Gang ``Rapper's Delight" (27);
      Evelyn Thomas ``Heartless" (29), Reflections (34)
       & 
      clipping distortion (63) \\
\hline
{\em Hip hop} &
      (39,45) (76,78) & (01,42) (46,65) (47,67) (48,68) (49,69) (50,72) & Wu-Tang Clan (1,7,41,42);
      Beastie Boys (8-25); A Tribe Called Quest (46-51,62-75); Cypress Hill (55-61); 
      Public Enemy (81-98); and others &  (02,32) & 
      3LW ``No More (Baby I'ma Do Right)'' (26); 
      Aaliyah ``Try Again'' (29); 
      Pink ``Can't Take Me Home" (31);
      Lauryn Hill ``Ex-Factor'' (40)
       & clipping distortion (3,5); skip at start (38)\\
\hline
{\em Jazz} &
      (33,51) (34,53) (35,55) (36,58)
(37,60) (38,62) (39,65) (40,67) (42,68) (43,69)
(44,70) (45,71) (46,72) & & James Carter (2-10); Joe Lovano (11-24); Branford Marsalis Trio (25-32);
Coleman Hawkins (33-46,51,53,55,57, 58,60,62,65,67-72); Dexter Gordon (73-77); Miles Davis (87-92); Joe Henderson (94-99); and others &   & 
Leonard Bernstein ``On the Town: Three Dance Episodes, Mvt. 1'' (00) and ``Symphonic dances from West Side Story, Prologue'' (01)
       & clipping distortion (52,54,66)\\
\hline
%Genre & \multicolumn{4}{c|@{\,}}{Repetitions} & \multicolumn{2}{c|@{\,}}{Mislabeling} & Distortion \\
% & Exact & Record. & Version & Artist & Wrong & Contentious & \\
%\hline
{\em Metal} & (04,13) (34,94) (40,61) (41,62) (42,63) (43,64) (44,65) (45,66) (58) is Rock (16)
      & & Dark Tranquillity (12-15); Dio (40-45,61-66); The New Bomb Turks (46-57); Queen (58-60);
      Metallica (33,38,73, 75,78,83,87); Iron Maiden (2,5,34,92-94); Rage Against the Machine (95-99); 
      and others &  (33,74) (85) is Ozzy Osbourne covering Disco (14) & 
      Creed ``I'm Eighteen'' (21); Living Colour ``Glamour Boys'' (29); 
      The New Bomb Turks ``Hammerless Nail'' (46), ``Jukebox Lean'' (50),
      ``Jeers of a Clown'' (51);
      Queen ``Tie Your Mother Down'' (58), ``Tear it up'' (59), ``We Will Rock You'' (60);
      Def Leppard ``Pour Some Sugar On Me'' (71), ``Photograph'' (79);
      Deep Purple ``Smoke On The Water'' (72);
      Bon Jovi ``You Give Love A Bad Name'' (86);
      Rage Against The Machine ``Wake Up'' (97)
       & clipping distortion (33,73,84)\\
\hline
{\em Pop} & (15,22) (30,31) (45,46) (47,80) (52,57) (54,60) (56,59) (67,71) (87,90)
      & (68,73) (15,21,22) (47,48,51) (52,54) (57,60) &  
      Mandy Moore (00,87-96); Mariah Carey (2,97-99); Alanis Morissette (3-9); Celine Dion (11,39,40); 
      Britney Spears (15-38); Christina Aguilera (44-51,80); Destiny's Child (52-62); 
      Janet Jackson (67-73); Jennifer Lopez (74-78,82); Madonna (84-86); and others &  
      (10,14) (16,17) (74,77) (75,82) (88,89) (93,94) &
      Alanis Morissette ``You Oughta Know'' (9);
      Destiny's Child ``Apple Pie a La Mode'' (61); 
      Diana Ross ``Ain't No Mountain High Enough'' (63);
      Prince ``The Beautiful Ones'' (65);
      Ladysmith Black Mambazo ``Leaning On The Everlasting Arm'' (81) 
      & (37) is from same recording as (15,21,22) but with sound effects \\
\hline
{\em Reggae} & (03,54) (05,56) (08,57) (10,60) (13,58) (41,69) (73,74) (80,81,82) (75,91,92)
      & (07,59) (33,44) (85,96) & Bob Marley (00-27,54-60); Dennis Brown (46-48,64-68,71); 
      Prince Buster (85,94-99); Burning Spear (33,42,44,50, 63); Gregory Isaacs (70,76-78); and others &  (23,55)  & 
      Pras ``Ghetto Supastar (That Is What You Are)'' (52);
      %Bounty Killer ``Hip-Hopera'' (73,74)
      & last 25 s of (86) are useless\\
\hline
{\em Rock} &  (16) is Metal (58)
      &  & Morphine (0-9); Ani DiFranco (10-15); Queen (16-26); The Rolling Stones (28-31,33,35,37);
      Led Zeppelin (32,39-48); Simple Minds (49-56); Sting (57-63); Jethro Tull (64-70); 
      Simply Red (71-78); Survivor (79-85); The Stone Roses (91-99) &  & 
      Morphine ``I Know You Pt. III'' (1), ``Early To Bed'' (2), ``Like Swimming'' (4);
      Ani DiFranco (10--15); 
      Queen ``(You're So Square) Baby I Don't Care'' (20);
      The Beach Boys ``Good Vibrations'' (27); 
      Billy Joel ``Movin' Out'' (36); 
      Guns N' Roses ``Knockin' On Heaven's Door'' (38);
      Led Zeppelin ``The Crunge'' (40), ``The Wanton Song'' (47);
      Sting ``Consider Me Gone'' (62), ``Moon Over Bourbon Street'' (63);
      Simply Red (71--78);
      The Tokens ``The Lion Sleeps Tonight'' (90)
        & jitter (27) \\
\hline
       \end{tabular}
%\vspace{-0.05in}
\caption{Repetitions, mislabelings and distortions in {\em GTZAN}. Excerpt numbers are in parentheses.}
%\vspace{-0.45in}
\label{tab:problems1}
\end{table*}

\restoregeometry

%%For instance, excerpts 98 and 99 in the category Disco
%%are the same portion of ``Rasputin'' by Boney M. 
%%As an example, Fig. \ref{fig:fingerprints} clearly shows 
%%13 exact repetitions in the excerpts of the category Jazz.
%We confirm each exact repetition by listening to the excerpts.
%Our comparison of fingerprints across categories
%reveal one exact repetition in two categories:
%the same excerpt of ``Tie Your Mother Down'' by Queen
%appears as Rock 58 and Metal 16.
%%one of which is in two different categories.

%\vspace{-0.15in}
To test for these conditions,
we use the scores between the 
top tag-normalized count pairs of the {\em GTZAN} labels,
\(C(\mathcal{T}_g,\mathcal{T}_r)\),
which we call {\em paired label scores}.
%(\ref{eq:score}) evaluated with the top tags and normalized counts
%of labels \(g\) and \(r\).
These values are shown in Table \ref{tab:scores}.
For example, the fourth element along the diagonal
is the score in {\em GTZAN} Disco
for a song having the same tags and normalized counts as
{\em GTZAN} Disco --- the ``perfect'' {\em GTZAN} Disco excerpt
since its tags reflect the majority description of {\em GTZAN} Disco.
The element to its right is its score in {\em GTZAN} Hip hop;
and the element to its left is its score in {\em GTZAN} Country.
We say an excerpt labeled \(g\) 
is mislabeled if its score in its label is an order of magnitude smaller
than the paired label score, i.e.,
\begin{equation}
C(\mathcal{X}, \mathcal{T}_g) < C(\mathcal{T}_g, \mathcal{T}_g)/10
%& < \max C(\mathcal{T}_r; \mathcal{X} \\
%\Delta ) := \sum_{j \in \mathcal{J}_r} \bar d_j(y_j) \sum_{i \in \mathcal{I}}   c_i  \delta_{y_j \equiv w_i}
\end{equation}
or if its score for another category \(r \ne g\) is too large, i.e.,
\begin{equation}
C(\mathcal{X}, \mathcal{T}_r) > C(\mathcal{T}_g, \mathcal{T}_g) - \Delta_g,
%& < \max C(\mathcal{T}_r; \mathcal{X} \\
% ) := \sum_{j \in \mathcal{J}_r} \bar d_j(y_j) \sum_{i \in \mathcal{I}}   c_i  \delta_{y_j \equiv w_i}
\end{equation}
where \(\Delta_g\) is one-tenth the largest magnitude difference between
all pairs of elements from \(\{C(\mathcal{T}_g, \mathcal{T}_r)\}_{g,r \in \mathcal{G}}\).
The values of \(\Delta_g\) for each label are shown in the last column of Table \ref{tab:scores}.
We divide by 10 in order to allow differences in scores 
that are an order of magnitude less than the largest difference in each row.
We also find experimentally that this finds many of the same mislabelings
%without too many false positives.
%Furthermore, this formalized definition of mislabeling
mentioned in our previous work \cite{Sturm2012b}.

%In the case of the Disco label, we see from Table \ref{tab:scores} that
%the maximum score occurs for the Disco label;
%but we also find overlap with
%the Pop and Rock labels --- which, as it reflects
%musicological underpinnings of all three genres,
%is evidence for the sanity of this approach to 
%detecting mislabelings in {\em GTZAN}.
%Furthermore, notice that $C(\mathcal{T}_g, \mathcal{T}_r) = C(\mathcal{T}_r, \mathcal{T}_g)$,
%and thus Table \ref{tab:scores} is symmetric.

%If the score for a tagged excerpt in category \(r\) is not larger than $C(r; \mathcal{Y}_r)/10$,
%then we consider it is potentially mislabeled.
For example, Pop 81 is ``Leaning On The Everlasting Arm'' by
Ladysmith Black Mambazo, and has top tags (with normalized counts):
``african'' (0.270), 	 ``world'' (0.195) and	 ``southafrica'' (0.097).
Its only non-zero score is $0.00010$ in {\em GTZAN} Rock,
and so we consider this excerpt mislabeled
(but not necessarily better labeled {\em GTZAN} Rock).
Disco 11 --- ``Can You Feel It'' by Billy Ocean from 1998 --- 
has top tags ``rock'' (0.33) and ``pop'' (0.33),
and has a score in {\em GTZAN} Disco of $0.018 > 0.0527/10$; 
but its score in {\em GTZAN} Pop is $0.06415 > 0.018 - 0.004$.
Hence, we consider it mislabeled.
%The highest overlap we see in this table 
%is between the Metal (Rock) category
%and the top tags of Rock (Metal).
We do not consider the 81 excerpts
we have yet to identify. % are mislabeled or not.
In total, we find 93 mislabelings,
and show 59 in Table \ref{tab:problems1}.
%The sixth column of Table \ref{tab:problems1} lists some of the 
%mislabelings we identify.
%ignoring false alarms due to uninformative tags ---
%e.g., Blues 1 is tagged only ``bonnie raitt''.
%For instance, the score for Country 17 (``San Antonio Rose'' by Floyd Cramer)
%in the top tags of Country is $0.00136 < 0.0947/10$.
%Its top tags are: ``instrumental'' (0.32), ``piano'' (0.18), and ``60s'' (0.11);
%but listening to it reveals the excerpt to not likely be mislabeled
%(the tag ``instrumental'' is not very genre-indicative).

\begin{table*}[t]
\tiny
\centering
\begin{tabular}{c|c|c|c|c|c|c|c|c|c|c||c|}
%\hline 
%& \multicolumn{10}{c||}{{\bf Excerpt having the same tags as top tags of {\em GTZAN} label}} & \\
& Blues & Classical & Country & Disco & Hip hop & Jazz & Metal & Pop & Reggae & Rock & $\Delta_g$ \\ \hline 
Blues & 0.0841 & 0 & 0 & 0 & 0 & 0 & 0 & 0 & 0 & 0 & 0.0084 \\ \hline 
Classical & 0 & 0.1261 & 0 & 0 & 0 & 0 & 0 & 0 & 0 & 0 & 0.0126 \\ \hline 
Country & 0 & 0 & 0.0947 & 0 & 0 & 0 & 0 & 0 & 0 & 0 & 0.0095 \\ \hline 
Disco & 0 & 0 & 0 & 0.0527 & 0.0008 & 0 & 0.0012 & 0.0124 & 0 & 0.0055 & 0.0040 \\ \hline 
Hip hop & 0 & 0 & 0 & 0.0008 & 0.0791 & 0 & 0.0004 & 0.0016 & 0 & 0.0014 & 0.0078 \\ \hline 
Jazz & 0 & 0 & 0 & 0 & 0 & 0.0830 & 0 & 0 & 0 & 0.0025 & 0.0081 \\ \hline 
Metal & 0 & 0 & 0 & 0.0012 & 0.0004 & 0 & 0.0367 & 0.0017 & 0 & 0.0158 & 0.0021 \\ \hline 
Pop & 0 & 0 & 0 & 0.0124 & 0.0016 & 0 & 0.0017 & 0.0453 & 0 & 0.0089 & 0.0033 \\ \hline 
Reggae & 0 & 0 & 0 & 0 & 0 & 0 & 0 & 0 & 0.1220 & 0 & 0.0122 \\ \hline 
Rock & 0 & 0 & 0 & 0.0055 & 0.0014 & 0.0025 & 0.0158 & 0.0089 & 0 & 0.0249 & 0.0009 \\ \hline 
\end{tabular}
\caption{The paired label scores for {\em GTZAN}, $C(\mathcal{T}_g,\mathcal{T}_r)$ in (\ref{eq:score}).
Last column shows values we use to test for significant differences.}
\label{tab:scores}
%\vspace{-0.1in}
\end{table*}

\begin{table*}[t]
\tiny
\centering
\begin{tabular}{c|c|c|c|c|c|c|c|c|c|c||c|}
%\hline 
& \multicolumn{10}{c||}{{\bf GTZAN label}} &  \\
& Blues & Classical & Country & Disco & Hip hop & Jazz & Metal & Pop & Reggae & Rock & Precision \\ \hline 
Blues       & 100 & 0     & 1      & 1         & 0          & 0       & 0        & 0.1 & 0            & 1  & 97.0 \\ \hline 
Classical & 0      & 100 & 0      & 0         & 0          & 2       & 0        & 0.1 & 0            & 0  & 97.9 \\ \hline 
Country   & 0      & 0     & 99 & 0        & 0          & 0       & 0        & 0.1 & 0            & 1   & 98.9 \\ \hline 
Disco       & 0      & 0     & 0     & 92       & 0          & 0       & 0        & 2.1 & 0            & 0   & 97.8 \\ \hline 
Hip hop   & 0      & 0     & 0      & 1        & 96.5     & 0       & 0        & 0.6 & 1            & 0  & 95.5 \\ \hline 
Jazz        & 0      & 0     & 0      & 0        & 0          & 98     & 0        & 0.1 & 0            & 4  & 96.0 \\ \hline 
Metal      & 0      & 0     & 0      & 0         & 0         & 0       & 90       & 0.1 & 0            & 4 & 95.6 \\ \hline 
Pop         & 0      & 0     & 0   & 5        & 3.5       & 0       & 0        & 95.6 & 0           & 15 & 79.9 \\ \hline 
Reggae   & 0      & 0     & 0      & 0        & 0          & 0       & 0        & 0.1  & 99          & 0  & 99.9 \\ \hline 
Rock        & 0      & 0    & 0      & 1        & 0          & 0       & 10       & 1.1 & 0            & 75  & 86.1 \\ \hline \hline
F-score& 98.5 & 98.9 & 98.9 & 94.8 & 95.8    & 97.0   & 92.7   & 87.0 & 99.4      & 80.2 &  Acc: 94.5\\ \hline
\end{tabular}
\caption{The statistics ($\times 10^{-2}$) of the ``perfect'' classifier for {\em GTZAN}
considering the 59 mislabelings in Table \ref{tab:problems1},
and that the 81 excerpts we have yet to identify have ``correct'' labels}
\label{tab:bestconfmatrix}
%\vspace{-0.35in}
\end{table*}

\subsection{Identifying faults: Distortions}
%\vspace{-0.05in}
The last column of Table \ref{tab:problems1}
lists some distortions we find by listening to every excerpt in {\em GTZAN}.
This dataset was purposely created to have
a variety of fidelities in the excerpts \cite{Tzanetakis2002};
however, one of the excerpts (Reggae 86) is so severely distorted 
that its last 25 seconds are useless.

%\vspace{-0.1in}
\subsection{Estimating statistics of ``perfect'' performance}
We now estimate several figures of merit for the ``perfect'' MGR system
evaluated using {\em Classify} \cite{Sturm2012d} in {\em GTZAN}.
Table \ref{tab:bestconfmatrix} estimates the ``ideal'' confusion table,
which we construct using the scores of the mislabeled excerpts in Table \ref{tab:problems1}.
For instance, %as described above in subsection \ref{sec:mislabelings},
Country 39 has its highest score in Blues,
so we add one to the Blues label in the Country column.
On the other hand, Hip hop 40 has its highest score in Hip hop,
but its score in Pop is high enough to consider it significant
(per the definition above), %in ection \ref{sec:mislabelings}),
so we add 0.5 to both Hip hop and Pop.
For Pop 81, where hardly any of the tags match
the top tags of the {\em GTZAN} labels,
we assign a weight of 0.1 to all labels.
We also assume that the first 5 seconds of Reggae 86
are representative enough of {\em GTZAN} Reggae.
Finally, we assume that the 81 excerpts we have yet to identify
have ``correct'' labels.
From Table \ref{tab:bestconfmatrix}, we can estimate the 
best recalls (the diagonal),
the best precisions (last column),
and the best F-scores (last row).
The best classification accuracy (bottom right corner) is around 94.5\%.
Hence, if the classification accuracy of
an MGR system tested in {\em GTZAN}
is better that this,
then it might actually be performing worse than the ``perfect'' system. % ---
%but classification accuracy is not enough to determine this \cite{Sturm2012e}.

%\vspace{-0.15in}
\section{{\em GTZAN} in MGR Research}
\begin{figure}[t]
\centering
\includegraphics[width=0.8\columnwidth]{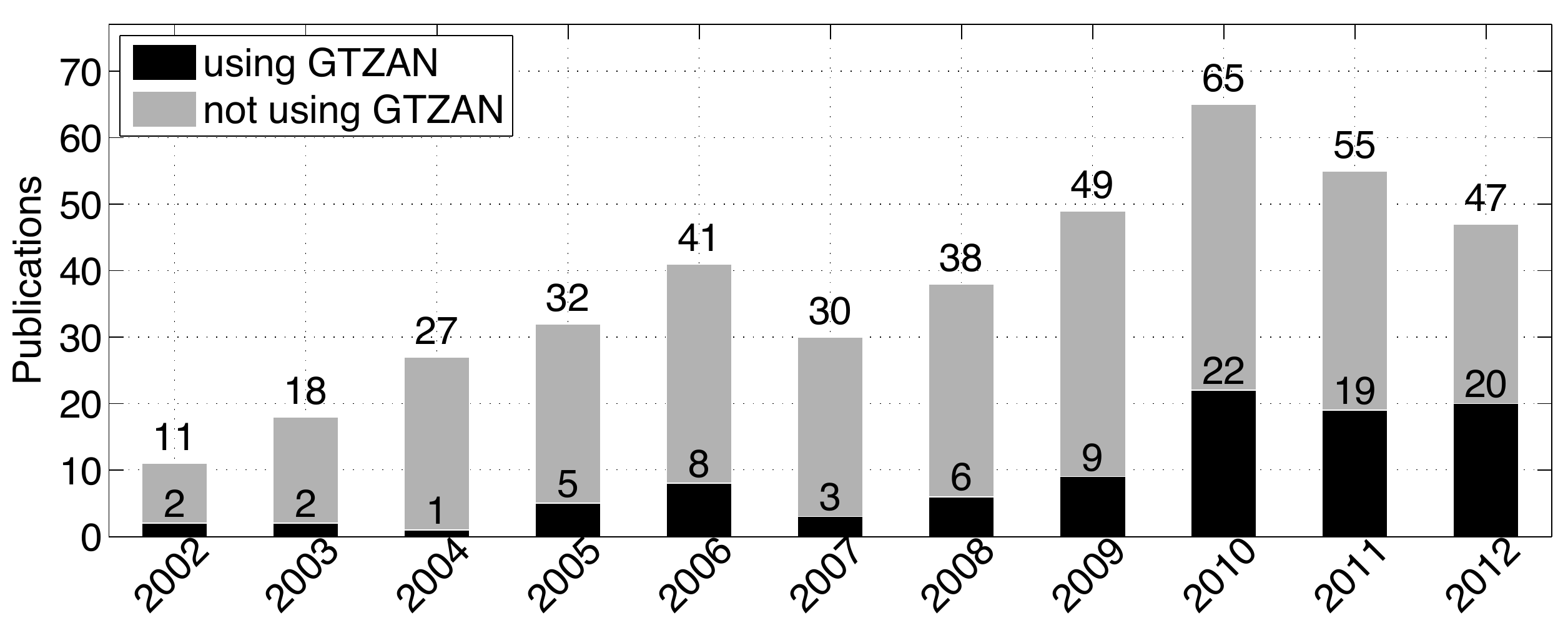}
%\vspace{-0.1in}
\caption{Annual numbers of published works in MGR 
with experimental components, divided into 
ones that use and do no use {\em GTZAN}.}
\label{fig:GTZANstats}
%\vspace{-0.15in}
\end{figure}

%To our knowledge, {\em GTZAN} is the first public MGR dataset,
%and has thus facilitated much research.
Figure \ref{fig:GTZANstats} shows how 
the number of publications that use {\em GTZAN}
has increased since its creation in 2002.
The next most used publicly-available dataset
is that created for the ISMIR 2004 MGR contest \cite{ISMIR2004},
which appears in 75 works,
30 of which use {\em GTZAN} as well \cite{Anglade2010,Dixon2010,
Holzapfel2007,Holzapfel2008,Lee2009b,Leon2012,
Leon2012b,Lidy2005,Lidy2006,Lidy2007,
Lidy2010b,Markov2012,Markov2012b,Mayer2010c,
Moerchen2006,Panagakis2008,Panagakis2009,Panagakis2009b,
Panagakis2010,Panagakis2010c,Ren2012,
Schindler2012b,Seo2011,Seyerlehner2010,Seyerlehner2010b,
Seyerlehner2012,Sturm2013b,Wu2011,Yeh2012,Zeng2009}.
Of the 100 works that use {\em GTZAN}, 49 of them only use {\em GTZAN}
\cite{Ahonen2010,Anden2011,Arabi2009,Bagci2007,Barreira2011,Behun2012,Benetos2008,Benetos2010,Bergstra2006b,Bergstra2010,Chang2010,Chathuranga2012,Chen2006,Chen2008,Draman2011,Fu2010,Fu2010b,Fu2011b,Genussov2010,Hamel2010,Henaff2011,Karkavitsas2011,Karkavitsas2012,Kotropoulos2010,Lampropoulos2010,Li2003b,Li2005,Li2010,Li2011,Lim2011,Manzagol2008,Nagathil2011,Ravelli2010,Ren2011,Rump2010,Santos2010,Serra2011,Shen2006,Sotiropoulos2008,Srinivasan2004,Sturm2012,Sturm2012c,Sturm2012e,Sturm2013,Turnbull2005,Wulfing2012,Yang2011b,Yaslan2006b,Zhou2012}.

%\vspace{-0.1in}
\subsection{Listening to {\em GTZAN}}
Among 100 works, 
we find only five (outside %of \cite{Tzanetakis2002,Tzanetakis2002b}
our recent work \cite{Sturm2012c,Sturm2012e,Sturm2013,Sturm2013b}),
that indicate someone has listened to at least some of {\em GTZAN}.
The first appears to be Li and Sleep \cite{Li2005}, 
who find that ``... two closely numbered files in each genre 
tend to sound similar than the files numbered far [apart].''
Bergstra et al. \cite{Bergstra2006} note that,
``To our ears, the examples are well-labeled ... %Although the artist names 
%are not associated with the songs, 
our impression from listening to the music 
is that no artist appears twice.''
This is contradicted by Seyerlehner et al. \cite{Seyerlehner2010b}, who predict 
``an artist effect ... as listening to some of the songs reveals that 
some artists are represented with several songs.''
In his doctoral dissertation, Seyerlehner \cite{Seyerlehner2010}
infers there to be a significant replication of artists
in {\em GTZAN} because of how classifiers trained and tested in that dataset perform
as compared to other artist-filtered datasets.

Very few works mention specific faults in {\em GTZAN}.
%we find some awareness. %authors must have listened to most, if not all, of the dataset.
Hartmann \cite{Hartmann2011} notes finding
seven duplicates, but mentions no specifics.
In \cite{Santos2010,Serra2011,Bogdanov2011,Marques2011c}, the authors describe
{\em GTZAN} as having 993 or 999 excerpts. %, while all other works describe it as having 1000.
In personal communication, de los Santos 
mentions that they found seven corrupted files in Classical   
(though \cite{Santos2010,Serra2011} report them being in Reggae).
%and \cite{Marques2011c} appears to say that {\em GTZAN} has 999 excerpts.
Li and Chan \cite{Li2011}, who
manually estimate the key of all {\em GTZAN} excerpts,
mention in personal communication that
they remember hearing some repetitions.
Their key estimates\footnote{http://visal.cs.cityu.edu.hk/downloads/\#gtzankeys}
are consistent among the exact repetitions we find in {\em GTZAN}.
%(http://visal.cs.cityu.edu.hk/downloads/\#gtzankeys).
%Finally, \cite{Lidy2006,Lidy2007} say the duration of the 
%dataset is 5h20, when 1000 30 s excerpts should be over 8h30.

%\vspace{-0.15in}
\subsection{Using {\em GTZAN}}
%\vspace{-0.05in}
In our review of evaluation in MGR \cite{Sturm2012d},
we delimit ten different experimental designs.
%as well as several figures of merit.
In the 100 works using {\em GTZAN},
96 employ the experimental design {\em Classify} 
\cite{Ahonen2010,Anden2011,Anglade2010,Arabi2009,Ariyaratne2012,Bagci2007,Barreira2011,Behun2012,Benetos2008,Benetos2010,Bergstra2006,Bergstra2006b,Bergstra2010,Chang2010,Chathuranga2012,Chen2006,Dixon2010,Draman2011,Fu2010,Fu2010b,Fu2011b,Genussov2010,Guaus2009,Hamel2010,Hartmann2011,Henaff2011,Holzapfel2007,Holzapfel2008,Karkavitsas2011,Karkavitsas2012,Kotropoulos2010,Krasser2012,Lampropoulos2010,Lee2009b,Leon2012,Leon2012b,Li2003,Li2003b,Li2005,Li2005b,Li2006,Li2010,Li2011,Lidy2005,Lidy2006,Lidy2007,Lidy2010b,Lim2011,Liu2009,Manzagol2008,Markov2012,Markov2012b,Marques2011c,Mayer2010c,Moerchen2006,Nagathil2011,Panagakis2008,Panagakis2009,Panagakis2009b,Panagakis2010,Panagakis2010c,Ravelli2010,Ren2011,Ren2012,Rocha2011,Rump2010,Salamon2012,Santos2010,Schindler2012b,Seo2011,Serra2011,Seyerlehner2010,Seyerlehner2010b,Shen2005,Shen2006,Sotiropoulos2008,Srinivasan2004,Sturm2012,Sturm2012c,Sturm2012e,Sturm2013,Sturm2013b,Tietche2012,Tsunoo2009,Tsunoo2009b,Tsunoo2011,Turnbull2005,Tzanetakis2002,Tzanetakis2002b,Wu2011,Wulfing2012,Yang2011b,Yaslan2006b,Yeh2012,Zeng2009,Yeh2013}
(an excerpt is assigned a class, 
and that class is compared against a ``ground truth'').
In seven papers, {\em GTZAN} is used with the design {\em Retrieve} 
\cite{Bogdanov2011,Chen2008,Fu2010,Leon2012,
Seyerlehner2012,Shen2006,Zhou2012}
(a query is used to find similar music,
and the labels of the retrieved items are compared).
The work in \cite{Barreira2011}
uses {\em GTZAN} in the design {\em Cluster} 
(data is clustered, and the composition of the resulting clusters are inspected).
Our work in \cite{Sturm2012c} uses {\em Compose}, 
where a system is trained on {\em GTZAN},
and then generates music it finds highly representative
of each {\em GTZAN} label.
With a formal listening test of these representative excerpts,
we find humans cannot recognize the genres they supposedly represent.

The design parameters in these works vary.
For {\em Classify},
most works measure MGR performance by
classification accuracy (the ratio of ``correct'' predictions to all observations) 
computed from
\(k\)-fold stratified cross-validation (\(k\)fCV),
e.g., \(2\)fCV (4 papers) \cite{Bagci2007,Fu2010,Fu2010b,Marques2011c}, 
\(3\)fCV (3 papers) \cite{Chen2006,Santos2010,Serra2011}, 
\(5\)fCV (6 papers) \cite{Anden2011,Bergstra2006b,Holzapfel2007,Holzapfel2008,Manzagol2008,Zeng2009}, 
 and \(10\)fCV (55 papers) \cite{Ahonen2010,Arabi2009,Behun2012,Benetos2010,Bergstra2010,Chang2010,Chathuranga2012,
Fu2011b,Genussov2010,Guaus2009,Hartmann2011,Henaff2011,Kotropoulos2010,Krasser2012,
Lee2009b,Leon2012b,Li2003,Li2003b,Li2005,Li2006,Lidy2005,Lidy2006,Lidy2007,Lidy2010b,Lim2011,
Mayer2010c,Moerchen2006,Panagakis2008,Panagakis2009,Panagakis2009b,Panagakis2010,Panagakis2010c,
Ren2011,Ren2012,Rocha2011,Salamon2012,Schindler2012b,Seo2011,Seyerlehner2010,Seyerlehner2010b,
Shen2005,Shen2006,Sturm2012,Sturm2012c,Sturm2012e,Sturm2013,Tsunoo2009,
Tsunoo2009b,Tsunoo2011,Turnbull2005,Wu2011,Wulfing2012,Yang2011b,Yeh2012,Yeh2013}.
Most of these use a single run of cross-validation;
however, some perform multiple runs, 
e.g., 10 independent runs of 2fCV (\(10\)x\(2\)CV) \cite{Marques2011c} or 20x2fCV \cite{Fu2010,Fu2010b},
10x3fCV \cite{Santos2010,Serra2011}, 
and 10x10fCV \cite{Lee2009b,Salamon2012,Schindler2012b,Seyerlehner2010,Sturm2012c,Sturm2012e,Sturm2013}.
In one experiment, Li and Sleep \cite{Li2005} use 10fCV
with random partitions;
but in another, they partition the excerpts
into folds based on their file number --- roughly implementing an
artist filter. %motivated by their observation that 
%files close in number sound like they come from the same artist.
Finally, leave one out cross-validation appears in \cite{Karkavitsas2011,Karkavitsas2012}.

Some works measure classification accuracy using a split of the data,
e.g., 60/40 \cite{Barreira2011}
(60\% used to train, 40\% used to test),
70/30 \cite{Benetos2008,Li2005b},
75/25 \cite{Nagathil2011},
80/20 \cite{Bergstra2006,Li2010,Liu2009,Ravelli2010},
90/10 \cite{Benetos2008,Srinivasan2004,Tzanetakis2002,Tzanetakis2002b,Yaslan2006b},
and training/validation/testing of 50/20/30 \cite{Hamel2010}.
Half of these report results from a single split 
\cite{Benetos2008,Barreira2011,Bergstra2006,Hamel2010,Li2010,Liu2009};
but the other half reports a mean of many trials,
e.g., 5 \cite{Li2005b}, 30 \cite{Yaslan2006b}, and 100 trials\cite{Ravelli2010,Srinivasan2004,Tzanetakis2002,Tzanetakis2002b}.
Seven papers \cite{Ariyaratne2012,Draman2011,Nagathil2011,Rump2010,Tietche2012,Leon2012,Marques2011c}
do not coherently describe their design parameters.

In some cases, only a portion of {\em GTZAN} is used, 
e.g., \cite{Anglade2010} performs 5x5fCV using only Classical, Jazz and Pop;
\cite{Lampropoulos2010} uses Blues, Classical, Country and Disco;
\cite{Sotiropoulos2008} uses several combinations of {\em GTZAN} categories;
and \cite{Dixon2010} states a three-genre subset is used, but provide no details 
on which they use.
Other works add to {\em GTZAN}, e.g.,
\cite{Barreira2011} adds 100 excerpts of Portuguese music
to {\em GTZAN} Classical, Metal, and Reggae;
\cite{Tietche2012} augments the excerpts in Classical and Jazz
with recordings of internet radio;
and \cite{Li2011} uses all of {\em GTZAN}, and augments it with
pitch-shifted and/or time-scaled versions.

\begin{landscape}
\begin{figure*}[t]
\centering
\includegraphics[width=1\columnwidth]{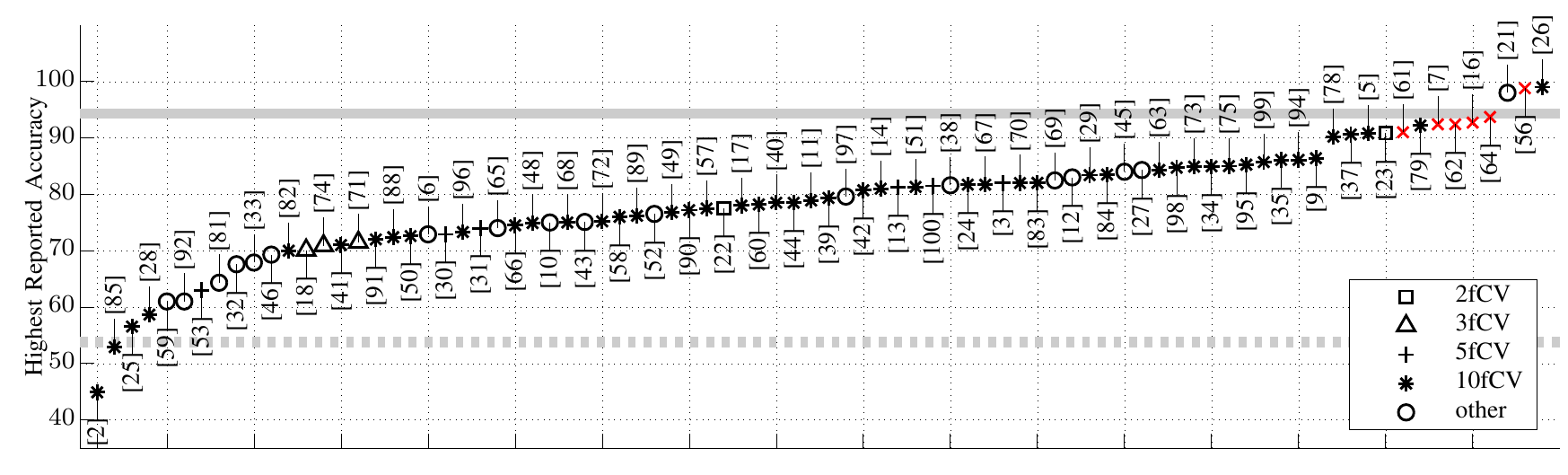}
%\vspace{-0.1in}
\caption{Highest classification accuracies (y-axis) reported
(cross-references labeled) with experimental design {\em Classify}
using all {\em GTZAN}.
Shapes (legend) denote particular details of the experimental procedure, 
e.g., ``2fCV'' is two-fold cross validation;
``other'' means randomly partitioning data into training/validation/test sets,
or an unspecified experimental procedure.
Five ``x'' denote results that have been challenged,
and/or shown to be invalid.
Solid gray line is our estimate of the ``perfect'' accuracy 
in Table \ref{tab:problems1}.
Dashed gray line is the best accuracy of the
five systems in Section \ref{sec:effects} that we evaluate using fault filtering.}
\label{fig:GTZANaccs}
%\vspace{-0.15in}
\end{figure*}
\end{landscape}

%In the {\em Retrieve} experimental design, 
%performance is most often measured by precision at \(k\), e.g.,
%\(k \inÊ\{5, 10, 15, 20\}\) \cite{Leon2012}; 
%\(k Ê= 10\) \cite{Seyerlehner2012};
%\(k Ê= 50\) \cite{Zhou2012}.
%Mean average precision is measured in \cite{Zhou2012,Bogdanov2011};
%and \cite{Fu2010} report average precision, as well as the
%receiver operating characteristics.
%Shen et al. \cite{Shen2006} report normalized precision and recall;
%and \cite{Chen2008} appears to use leave-one-out cross validation to compute recall.
%In the experimental design {\em Cluster} occurring in \cite{Barreira2011},
%the authors look at the composition of the resulting clusters.
%to find differences between labels.
%They find some feature combinations give rise to clusters
%that consist of excerpts having one {\em GTZAN} label.

%Finally, we find a few other uses of {\em GTZAN}.
%Markov et al. \cite{Markov2012,Markov2012b} 
%learn bases from {\em GTZAN}, and then apply these
%codebooks to classify the genres used by music in the {\em ISMIR2004} dataset.
%In \cite{Sturm2013b}, we train classifiers using {\em ISMIR2004}, 
%and then attempt to detect all excerpts in {\em GTZAN}
%that use the classical genre. %(100 in Classical, and 2 in Jazz).

In summary, we find that about half of the work
using {\em GTZAN} uses no other dataset,
which means that a majority of evaluations
can provide no conclusions about the performance of a system
on other datasets, e.g., in the real world \cite{Urbano2011b,Sturm2013e}.
Since 96 of 100 papers use {\em Classify}  
to evaluate MGR systems,
most papers report classification accuracy as a figure of merit,
and some work use this figure to compare systems
for recognizing music genre.
We have criticized this approach to evaluation \cite{Sturm2012e},
and argued that it lacks the validity necessary to make any meaningful comparisons
between systems \cite{Sturm2013e}.

%\vspace{-0.15in}
\subsection{Reported classification accuracies in {\em GTZAN}}
Figure \ref{fig:GTZANaccs} shows the highest
classification accuracies %(mean recall across the classes) 
reported in 96 papers that consider the 10-class problem of {\em GTZAN}
(we remove duplicated experiments,
e.g., \cite{Lidy2006} contains the results reported in \cite{Lidy2005}).
The thick gray line in Fig. \ref{fig:GTZANaccs} 
shows our estimate of the ``perfect'' classification accuracy from Table \ref{tab:bestconfmatrix}.
%We see most of the results are between 70-86\%.
Six results (marked with a red ``x'') are incorrect
or have been challenged:
the results in \cite{Panagakis2009,Panagakis2009b,Panagakis2010c}
are due to a mistake in the experimental procedure \cite{Sturm2012e},
as are those in \cite{Marques2011c};\footnote{Personal communication with J. P. Papa.}
the result in \cite{Chang2010} contradicts many others \cite{Sturm2013};
and the results in \cite{Bagci2007}
are not likely to come from the system \cite{Sturm2013d}.
\section{The faults of {\em GTZAN} and evaluation}\label{sec:effects}
In this section, we disprove the following two claims:
1) ``all MGR systems and evaluations
are affected in the same ways by the faults in {\em GTZAN}''; 
% compare system across conditions
and 2) ``the performances of all MGR systems in {\em GTZAN}, 
working with the same data and faults, are still meaningfully comparable.''
Such claims have been made in reviews of this work, 
but also appear in \cite{Yeh2013}.
% compare condition across systems
We now study how the faults of {\em GTZAN} 
affect the evaluation of MGR systems,
e.g., the estimation of classification accuracy 
using {\em Classify} in {\em GTZAN}. 

It is not difficult to predict how these faults
can affect the evaluations of particular systems.
For instance, when exact replicas 
are distributed across train and test sets,
the evaluation of some systems can be more biased than others: 
a nearest neighbor classifier
will find features in the training set with zero distance to the test feature,
while a Bayesian classifier with a parametric model
may not so strongly benefit when 
its model parameters are estimated
from all training features.
If there are replicas in the test set only,
then they will bias our estimate of a figure of merit
because they are not independent tests ---
if one is classified (in)correctly then its replicas are also classified (in)correctly.
Thus, we already have a notion that the two claims above 
are not true.

In addition to exact repetitions, we show above that 
the number of artists in {\em GTZAN} is at most 329.
Thus, as Seyerlehner \cite{Seyerlehner2010,Seyerlehner2010b} predicts for {\em GTZAN},
the evaluation of systems will certainly be biased
due to the {\em artist effect} \cite{Pampalk2005b,Flexer2007,Flexer2009,Flexer2010},
i.e., the observation that a music similarity system can perform significantly worse
when artists are disjoint in training and test datasets,
than when they are not.
Since {\em all} results in Fig. \ref{fig:GTZANaccs} come from
evaluations without an artist filter,
they will likely be optimistic.
What has yet to be shown for {\em GTZAN}, however,
is just how optimistic they might be.
After presenting our experimental method,
we then present our results,
and discuss the veracity of the two claims above.

%\vspace{-0.15in}
\subsection{Method}
%We use {\em GTZAN} based on the scope of these claims.
We use three classifiers with the same features \cite{Theodoridis2009}:
nearest neighbor (NN);
minimum distance (MD);
and minimum Mahalanobis distance (MMD).
We implement these classifiers in PRTools \cite{PRTools}.
We create feature vectors from a 30-s excerpt 
in the following way.
For each 46.4 ms frame, and a hop half that, we compute: 
13 MFCCs using the approach in \cite{Slaney1998},
zero crossings, and spectral centroid and rolloff.
For each 130 consecutive frames,
we compute the mean and variance of each dimension,
thus producing nine 32-dimensional feature vectors for each excerpt.
We normalize the dimensions of the training set features,
i.e., we find and apply the transformation mapping each dimension to \([0,1]\).
We apply the same transformation to the test set.
Each classifier labels an excerpt as follows:
NN randomly selects among the majority labels given to the nine feature vectors;
MD and MMD both select the label with the 
maximum log posterior sum over the nine feature vectors.

%Since one can argue these systems are elementary
%and so do not reflect real approaches to MGR,
In addition to there, we test two state-of-the-art MGR systems that produce some of the highest
classification accuracies reported in {\em GTZAN}.
The first is SRCAM --- proposed in \cite{Panagakis2009} and modified in \cite{Sturm2012e} ---
which uses psychoacoustically-motivated features in 768 dimensions.
%Each excerpt in {\em GTZAN} produces one of these  features.
SRCAM classifies an excerpt by sparse representation classification \cite{Wright2009b}.
We use the SPGL1 solver \cite{Berg2008}
with at most 200 iterations, and \(\epsilon^2 := 0.01\).
%(See \cite{Sturm2012e} for details.)
The second system is MAPsCAT, 
which uses features computed with the scattering transform \cite{Anden2011}.
This produces 40 feature vectors of 469 dimensions for a 30-s excerpt.
MAPsCAT estimates from the training set 
the mean for each class, the total covariance matrix,
and computes for a test feature the log posterior in each class.
We define all classes equally likely for MAPsCAT, as well as MD and MMD.
We normalize the features of the training and test sets.
We give further details of SRCAM and MAPsCAT in \cite{Sturm2012e}.

\begin{table*}[t]
\centering
\scriptsize
\begin{tabular}{@{\;}c@{\;}|@{\;}p{2.2in}@{\;}|@{\;}p{2.2in}@{\;}|}
%\hline 
& \multicolumn{1}{c|@{\;}}{{\bf Fold 1}} & \multicolumn{1}{c|}{{\bf Fold 2}} \\ \hline 
Blues       & John Lee Hooker, Kelly Joe Phelps, Buckwheat Zydeco, Magic Slim \& The Teardrops & 
Robert Johnson, Stevie Ray Vaughan, Clifton Chenier, Hot Toddy, Albert Collins \\ \hline
Classical & J. S. Bach, Percy Grainger, Maurice Ravel, Henri Dutilleux, Tchaikovsky, Franz Schubert, {\em Leonard Bernstein}, misc. & Beethoven, Franz Joseph Haydn, Mozart, Vivaldi, Claude Debussy, misc. \\ \hline
Country   &  Shania Twain, Johnny Cash, Willie Nelson, misc. & Brad Paisley, George Strait, Vince Gill, misc. \\ \hline 
Disco       & Donna Summer, KC and The Sunshine Band, Ottawan, The Gibson Brothers, Heatwave, Evelyn Thomas,  misc. & Carl Douglas, Village People, The Trammps, Earth Wind and Fire, Boney M., ABBA, Gloria Gaynor, misc. \\ \hline 
Hip hop   & De La Soul, Ice Cube, Wu-Tang Clan, Cypress Hill, Beastie Boys, 50 Cent, Eminem, misc. &
A Tribe Called Quest, Public Enemy, {\em Lauryn Hill}, Wyclef Jean \\ \hline 
Jazz        & {\em Leonard Bernstein}, Coleman Hawkins, Branford Marsalis Trio, misc. & 
James Carter, Joe Lovano, Dexter Gordon, Tony Williams, Miles Davis, Joe Henderson, misc.\\ \hline 
Metal      & Judas Priest, Black Sabbath, {\em Queen}, Dio, Def Leppard, Rage Against the Machine, {\em Guns N' Roses}, New Bomb Turks, misc. & 
AC/DC, Dark Tranquillity, Iron Maiden, Ozzy Osbourne, Metallica, misc. \\ \hline 
Pop         & Mariah Carey, Celine Dion, Britney Spears, Alanis Morissette, Christina Aguilera, misc. &
Destiny's Child, Mandy Moore, Jennifer Lopez, Janet Jackson, Madonna, misc. \\ \hline 
Reggae   & Burning Spear, Desmond Dekker, Jimmy Cliff, Bounty Killer, Dennis Brown, Gregory Isaacs, Ini Kamoze, misc. & 
Peter Tosh, Prince Buster, Bob Marley, {\em Lauryn Hill}, misc. \\ \hline 
Rock        & Sting, Simply Red, {\em Queen}, Survivor, {\em Guns N' Roses}, The Stone Roses, misc. &
The Rolling Stones, Ani DiFranco, Led Zeppelin, Simple Minds, Morphine, misc. \\ \hline
\end{tabular}
\caption{Composition of each fold of the artist filter partitioning (500 excerpts in each).
Italicized artists appear in two categories.}
\label{tab:foldcomposition}
%\vspace{-0.15in}
\end{table*}

We evaluate each system using {\em GTZAN} with four different kinds of partitioning:
ten realizations of standard non-stratified 2fCV (ST);
ST without the 68 exact and recording repetitions and 2 distortions (ST');
a non-stratified 2fCV with artist filtering (AF);
AF without the 68 exact and recording repetitions and 2 distortions (AF').
Table \ref{tab:foldcomposition} shows the composition of each fold of AF
in terms of artists.
We created AF manually to ensure that: 1) each class is approximately balanced 
in terms of the number of training and testing excerpts; and 2)
each fold of a class has music
representative of its top tags (Fig. \ref{fig:GTZANtags}).
For instance, the Blues folds have 46 and 54 excerpts,
and both have Delta blues and zydeco.
We retain the original labels of all excerpts,
and evaluate all systems using the same partitions.
%In every test, we use every excerpt only once.
%We make no attempt in our random partitioning of the dataset 
%to ensure repetitions are evenly split across training and test sets.
%Since we show above that 95 of 99 works evaluating MGR systems using {\em GTZAN}
%use classification accuracy as a figure of merit, we do the same.

We look at several figures of merit:
confusion, precision, recall, F-score,
and classification accuracy.
Consider that the \(i\)th fold of a cross-validation experiment
consists of \(N_{i}^{(g)}\)
excerpts of label \(g \in \mathcal{G}\),
and that of these a system classifies as \(\textrm{r} \in \mathcal{G}\) 
the number \(M_{i}^{(g \textrm{ as } r)} \le N_{i}^{(\textrm{g})}\).
We define the confusion of label \(g \textrm{ as } r\) of this fold as 
\begin{equation}
C_{i}^{(g \textrm{ as } r)} := M_{i}^{(g \textrm{ as } r)}\bigr /N_{i}^{(g)}.
\label{eq:confusion}
\end{equation}
Thus, $C_{i}^{(g \textrm{ as } g)}$ is the recall for class $g$ in the $i$th fold.
The normalized accuracy of a system in the $i$th fold is defined by
\begin{equation}
A_i := \frac{1}{|\mathcal{G}|}\sum_{g \in \mathcal{G}} C_i^{(g \textrm{ as } g)}.
\label{eq:accuracy}
\end{equation}
We use normalized accuracy because the classes
are not equally represented in the test sets.
Finally, the precision of a system for class X in the $i$th fold is
\begin{equation}
P_i^{(\textrm{X})} := M_i^{(\textrm{X as X})}\bigr / \sum_{Y\in\mathcal{G}} M_i^{(\textrm{Y as X})}
\label{eq:precision}
\end{equation}
and the F-score of an system for label X in the $i$th fold is
\begin{equation}
F_i^{(\textrm{X})} := 2 P_i^{(\textrm{X})} C_i^{(\textrm{X as X})} \bigr / \left [ P_i^{(\textrm{X})} + C_i^{(\textrm{X as X})} \right ].
\label{eq:Fscore}
\end{equation}

To test for significant differences in performance between two systems,
we first build contingency tables for all observations they classify \cite{Salzberg1997}.
Define the rv $N$ to be the number of times the two systems
choose different classes, but one is correct.
Let \(t_{12}\) be the number for which
system 1 is correct but system 2 is wrong.
Thus, \(N-t_{12}\) is the number of observations
for which system 2 is correct but system 1 is wrong.
Define the rv $T_{12}$
from which \(t_{12}\) is a sample.
The null hypothesis is that the systems perform equally well given $N=n$,
i.e., $E[T_{12}|N=n] = n/2$,
in which case $T_{12}$ is distributed binomial, i.e.,
\begin{equation}
p_{T_{12}|N=n}(t) = {n \choose t}(0.5)^{n}, 0 \le t \le n.
\end{equation}
The probability we observe a particular performance
given the systems perform equally well is %\vspace{-0.1in}
\begin{multline}
p := P[T_{12} \le \min(t_{12},n-t_{12})] + P[T_{12} \ge \max(t_{12},n-t_{12})] \\
= \sum_{t=0}^{\min(t_{12},n-t_{12})} p_{T_{12}|N=n}(t) 
+ \sum_{t=\max(t_{12},n-t_{12})}^{n} p_{T_{12}|N=n}(t).
\end{multline}
%\begin{equation}
%p := P[T_{12} \ge t_{12} | q = 0.5]
%= P\left [ \chi^2 \ge \frac{(|t_{12} - t_{21}|-1)^2}{t_{12} + t_{21}} \right ]
%\end{equation}
%when $t_{12} + t_{21}$ is large ($> 25$).
We define statistical significance as $\alpha = 0.05$.
In other words, we reject the null hypothesis
when $p < \alpha$.

\begin{figure}[t]
%\vspace{-0.25in}
\centering
\includegraphics[width=0.8\columnwidth]{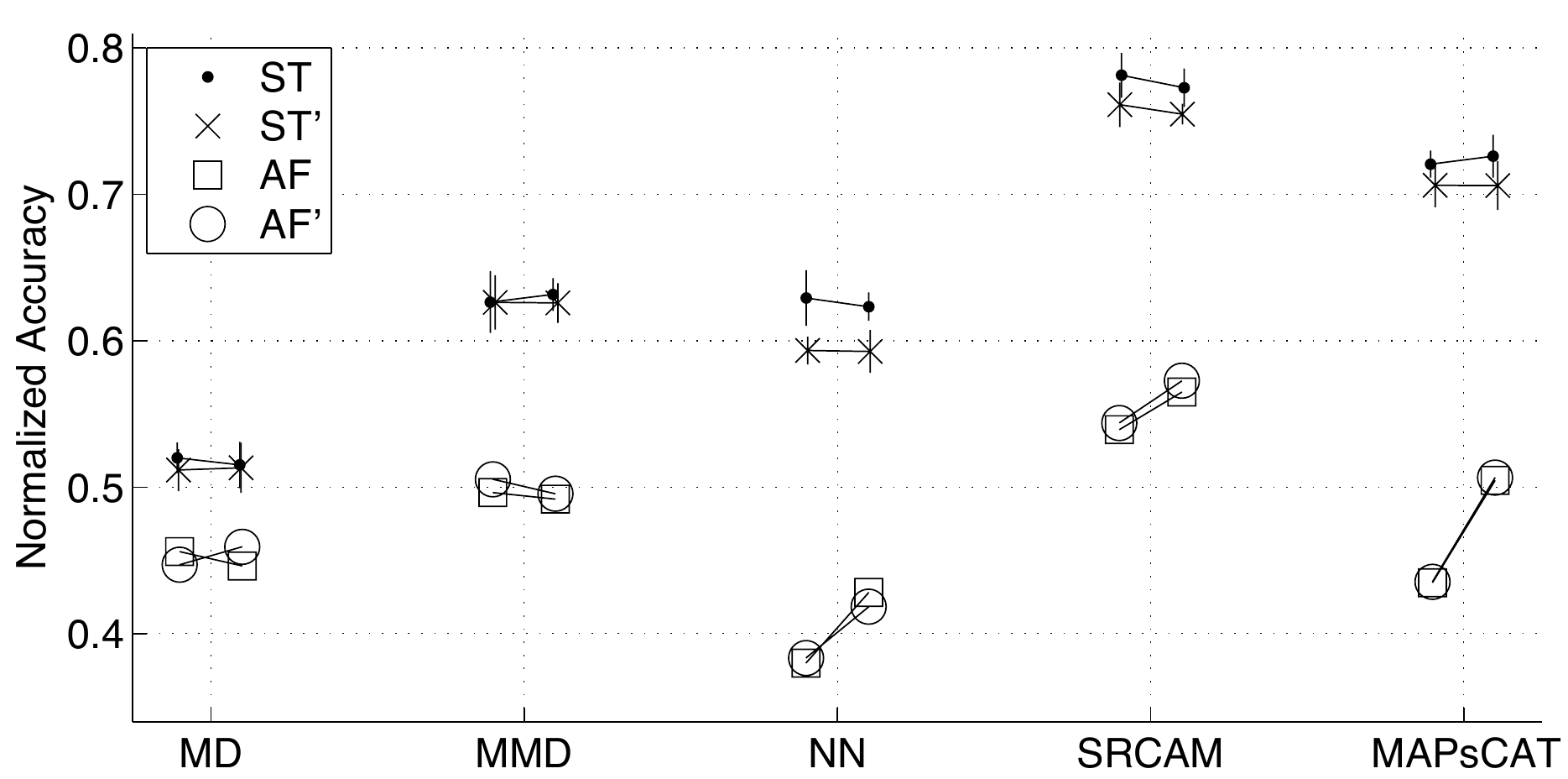}
\caption{Normalized accuracy (\ref{eq:accuracy}) of each system (x-axis)
for each fold (left and right) of different partitioning (legend).}
\label{fig:accuracies}
%\vspace{-0.15in}
\end{figure}

%\vspace{-0.15in}
\subsection{Experimental results and discussion}
Figure \ref{fig:accuracies} shows the normalized accuracies (\ref{eq:accuracy})
of all five systems for each fold of the four different kinds of partitioning.
For the ten partitions of ST and ST', we show the mean normalized accuracy,
and one standard deviation above and below.
It is immediately clear that estimates of the
classification accuracy for each system are affected by the faults of {\em GTZAN}.
We see that the differences between ST and ST' are small
for all systems except NN.
As we predict above, the performance of NN 
appears to benefit more than the others from the
exact and recording repetitions, which boost its 
mean normalized accuracy from about 0.6 to 0.63.
Between AF and AF', removing the repetitions produces very little change
since the artist filter keeps exact and recording repetitions from 
appearing in both train and test sets.
Most clearly, we see for all five systems large decreases in performance  
between ST and AF.
The difference in normalized accuracy of MD 
between ST and AF appears the smallest (7 points),
while that of MAPsCAT appears the most (25 points).
Since we have thus found systems
with performance evaluations affected to different magnitudes
by the faults in {\em GTZAN} --- that of NN is hurt
by removing the repetitions while that of MMD is not ---
this disproves the first claim above.

Testing for significant difference in performance
between all pairs of system in both ST and ST',
only for MMD and NN do we fail to reject the null hypothesis.
Furthermore, in terms of classification accuracy in ST, 
we can say with statistical significance:
MD $< \{$MMD, NN$\} <$ MAPsCAT $<$ SRCAM,
i.e., SRCAM performs the best and MD performs the worst.
In both AF and AF', however,
for MAPsCAT and MD, and for MAPsCAT and MMD,
do we fail to reject the null hypothesis.
In this case, 
we can say with statistical significance:
NN $<\{$MAPsCAT, MD$\} < \{$MAPsCAT, MMD$\} <$ SRCAM.
Therefore, while our conclusion on the basis of
our evaluation in ST is that MAPsCAT performs significantly better 
than all these systems except SRCAM,
its evaluation in AF says otherwise.
This disproves the second claim above.

\begin{figure*}[t]
%\vspace{-0.2in}
\centering
%\subfigure[MAPsCAT]{
%\includegraphics[height=3.35in]{figures/GTZANresults/class_Scattering/confusion_MAPsCAT_artist}}\hspace{-0.32in}
\subfigure[SRCAM, ST]{
\includegraphics[height=3in]{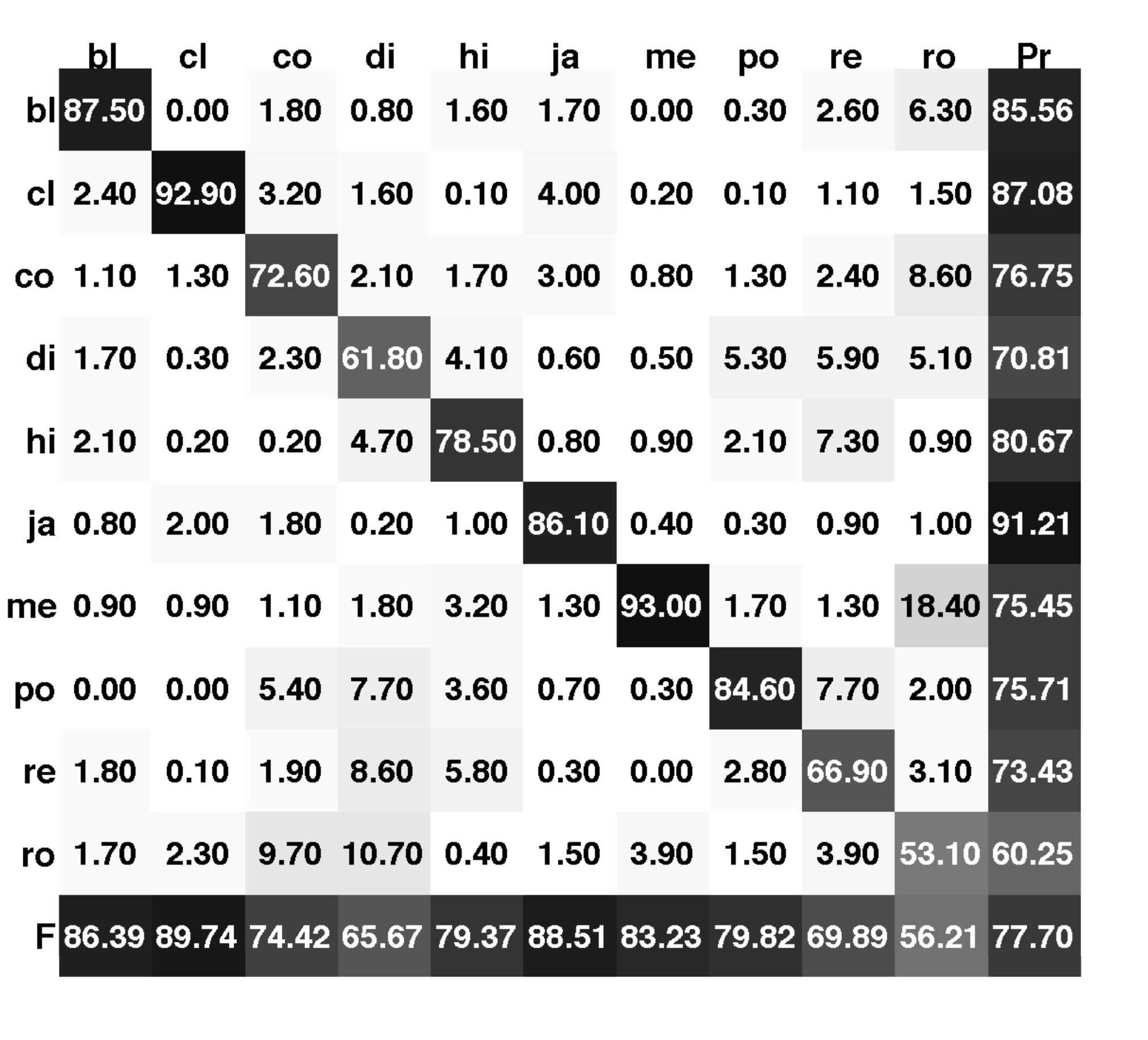}}\hspace{-0.30in}
\subfigure[SRCAM, ST']{
\includegraphics[height=3in]{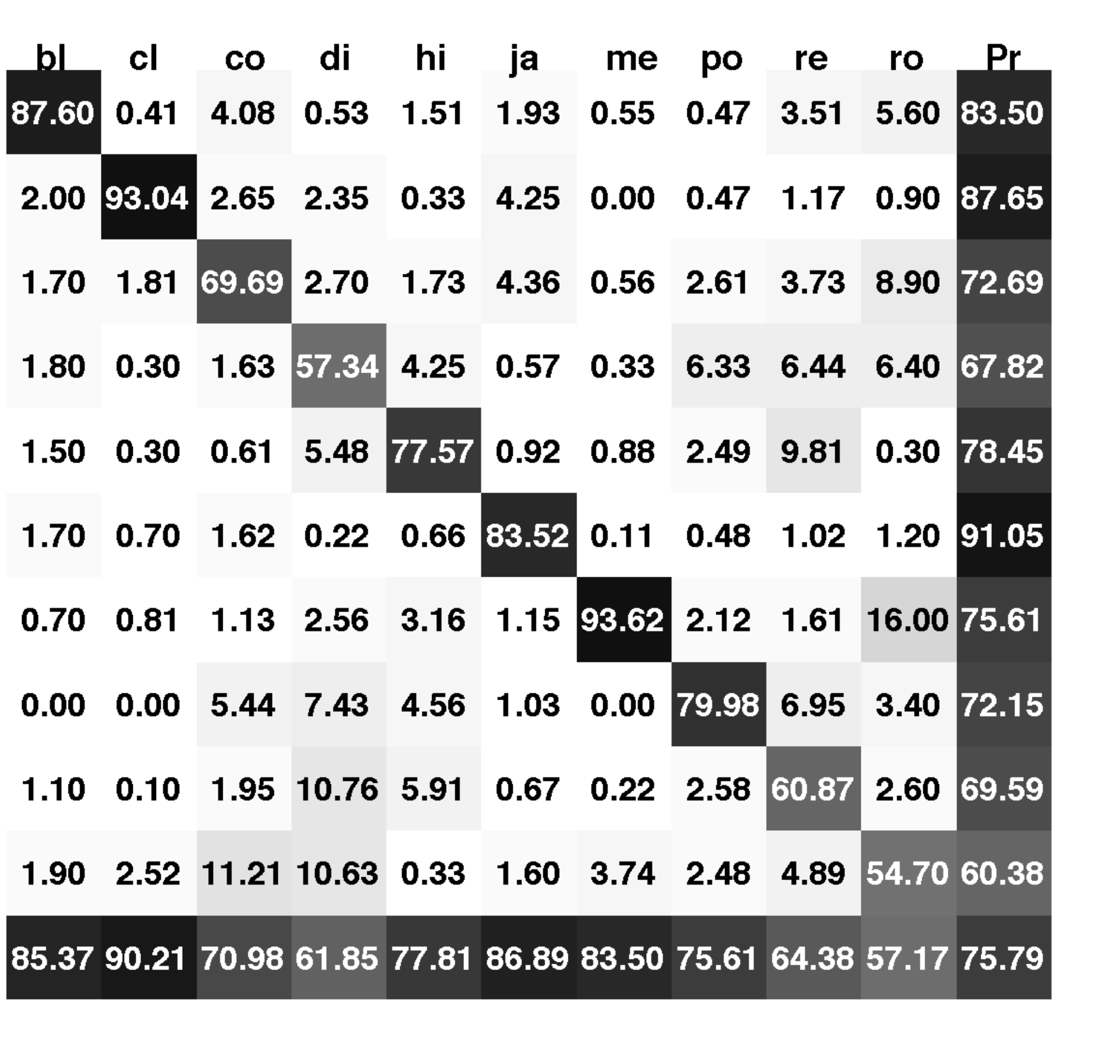}}
%\subfigure[SRCAM, AF]{
%\includegraphics[height=2.9in]{figures/GTZANresults/class_ATM/confusion_SRCAM_artist}}\hspace{-0.32in}
\subfigure[SRCAM, AF']{
\includegraphics[height=3in]{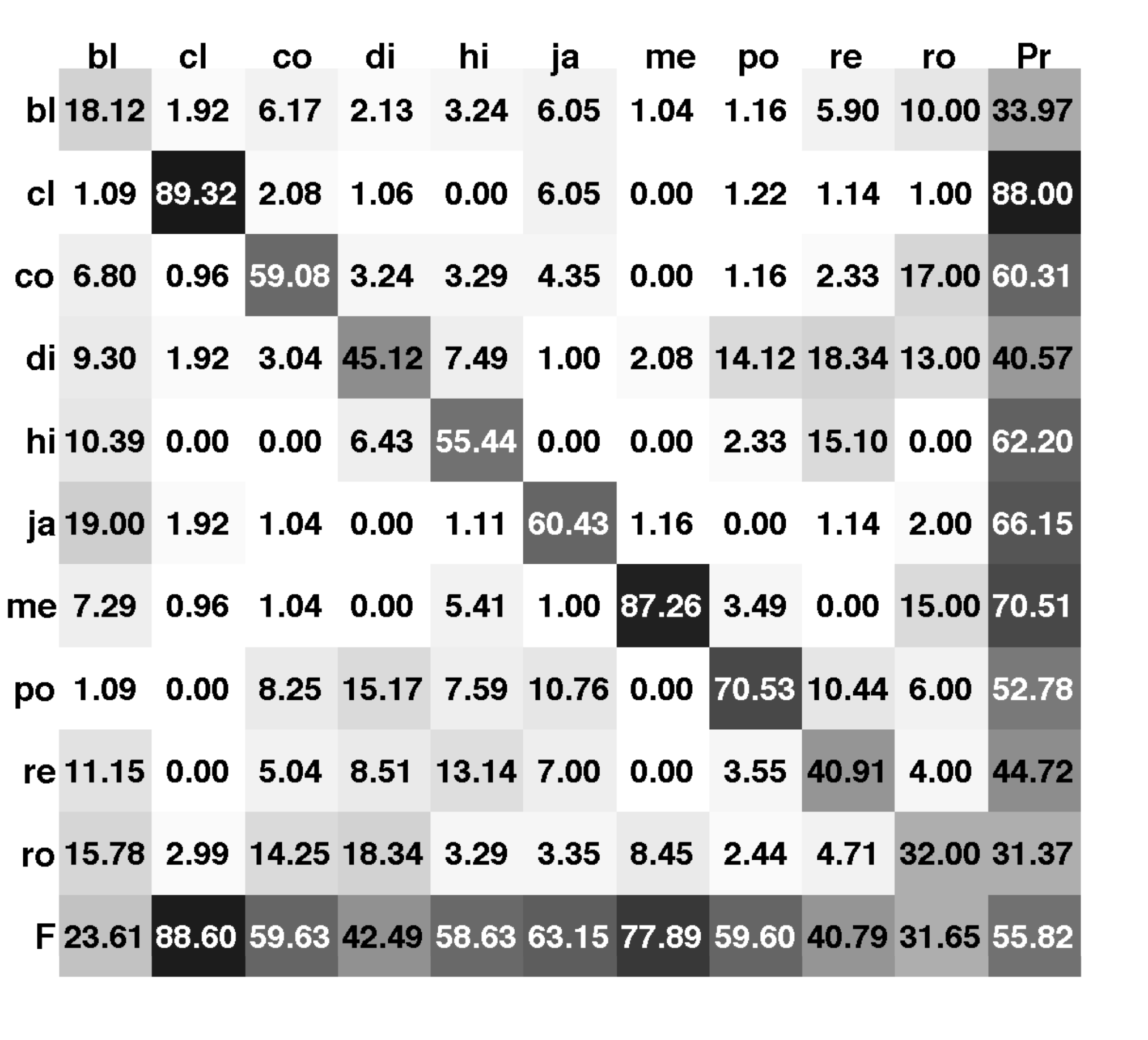}}\hspace{-0.30in}
\subfigure[SRCAM, AF' with relabeling using (\ref{eq:score})]{
\includegraphics[height=3in]{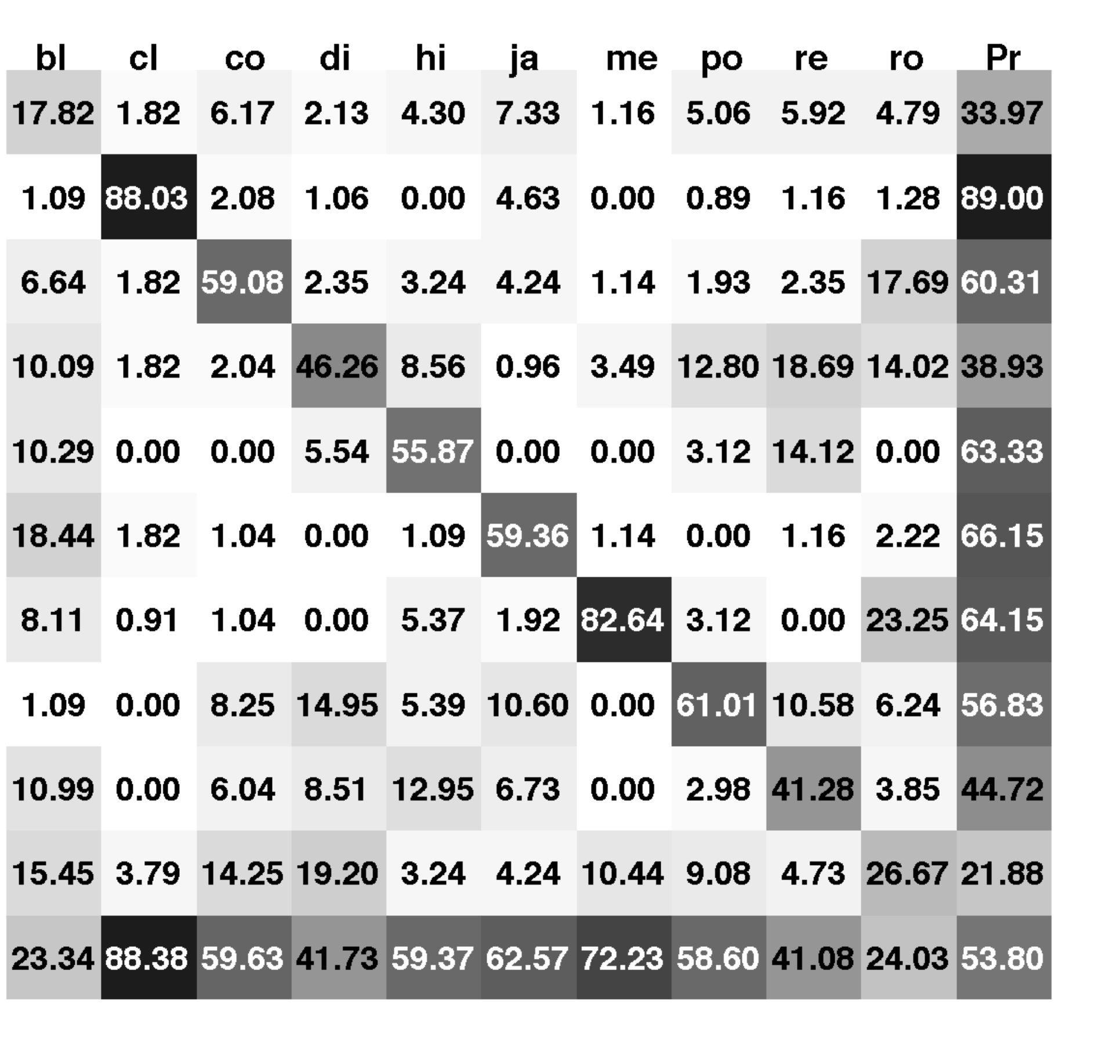}}
%\subfigure[MAPsCAT, ST]{
%\includegraphics[height=3.05in]{figures/GTZANresults/class_Scattering/confusion_MAPsCAT_none}}
%\vspace{-0.05in}
\caption{Confusion, precision (Pr), F-score (F),
and normalized accuracy (bottom right corner) for SRCAM 
evaluated with (ST) and without repetitions (ST')
averaged over 10 realizations,
as well as with artist filtering and without replicas (AF'),
and finally taking mislabelings into account.
Columns are ``true'' labels;
rows are predictions.
Darkness of square corresponds to value.
Labels: Blues (bl), Classical (cl), Country (co),
Disco (di), Hip hop (hi), Jazz (ja), Metal (me),
Pop (po), Reggae (re), Rock (ro).}
\label{fig:confusions}
%\vspace{-0.15in}
\end{figure*}

\begin{figure}[t]
%\vspace{-0.2in}
\centering
\includegraphics[width=0.8\columnwidth]{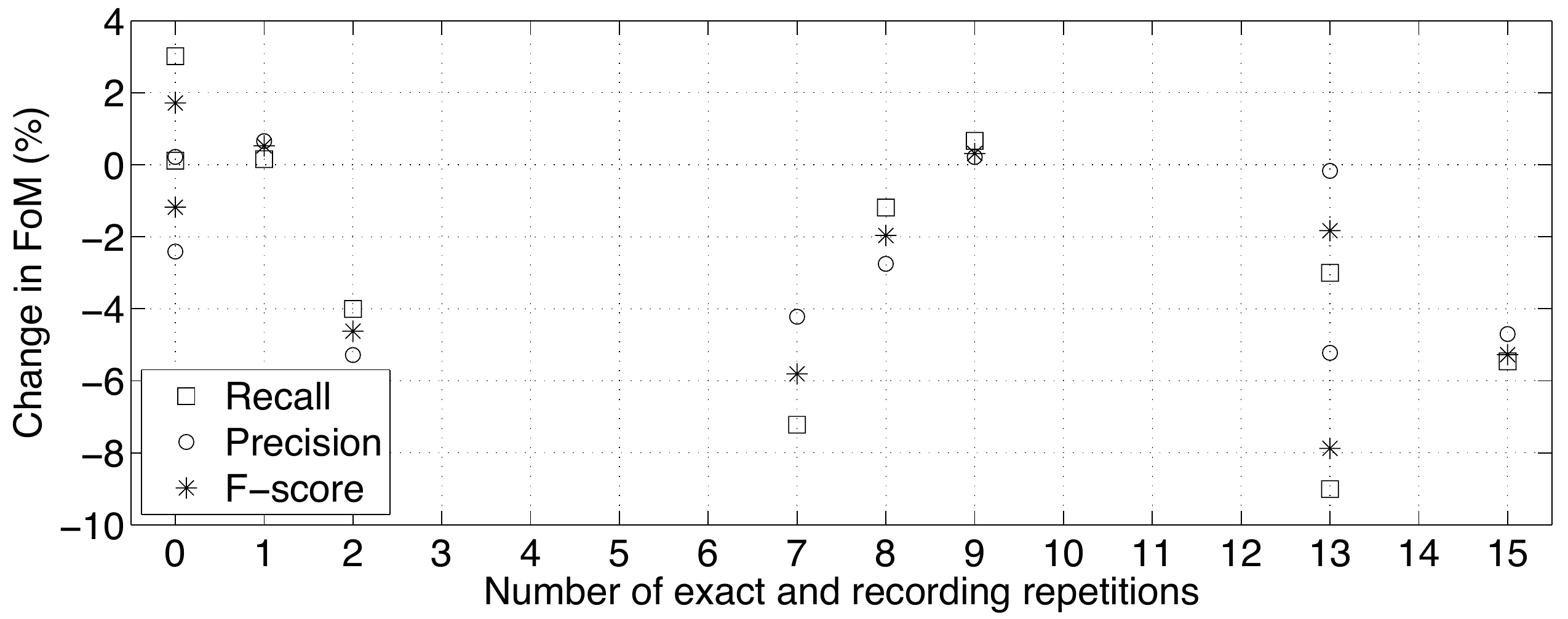}
%\vspace{-0.2in}
\caption{Scatter plot of percent change in mean figures of merit (FoM)
in Fig. \ref{fig:confusions} between ST and ST'
as a function of the number of exact and recording
repetitions in a class.}
\label{fig:scatter}
%\vspace{-0.2in}
\end{figure}

We now focus our analysis upon SRCAM.
Figure \ref{fig:confusions} shows other figures of merit 
for SRCAM averaged over 10 realizations of ST and ST' partitions, 
as well as for the single partition AF'.
Between ST and ST',
we see very little change in the recalls for classes
with the few exact and recording repetitions:
Blues, Classical and Rock.
However, for the classes having the most
exact and recording repetitions, we find large changes in their recalls.
In fact, Fig. \ref{fig:scatter} shows that
the number of exact and recording repetitions in a category
is correlated with a decrease in its recall for SRCAM.
This makes sense as SRCAM can be seen as a kind of adaptive nearest neighbors. 

When evaluating SRCAM in AF',
Fig. \ref{fig:confusions} shows our estimate of its classification accuracy in ST
decreases by 22 points (28\%).
With respect to F-score,
Classical appears to suffers the least;
but we see decreases for all other classes by at least 10\%, e.g.,
77\% for Blues, 46\% for Rock, and 44\% for Reggae.
We see little change in our estimated classification accuracy
when testing in AF' instead of AF (not shown), 
but our estimates of recalls, precisions, and F-scores for four classes increase,
while those for thee classes decrease.

In all figures of merit above, we have not considered the 59 mislabelings in {\em GTZAN}
noted in Table \ref{tab:problems1}.
We thus assign new labels to each of the 59 excerpts
based on which class has highest score (\ref{eq:score})
for the excerpt.
If the score is zero in every class, e.g., Pop 81, 
we keep the original label.
We assume that the 81 excerpts we have yet to identify
have ``correct'' labels.
The new figures of merit for SRCAM, seen in Fig. \ref{fig:confusions}(d),
show even further deterioration.
Compared to the figures of merit in ST,
we are quite far from the ``perfect'' statistics
in Table \ref{tab:bestconfmatrix}.

\newgeometry{top=1in, bottom=1in, left=0.6in, right=0.6in}
\restoregeometry

Based on these results, one might be inclined to argue that,
though it is clearly performing poorly overall,
SRCAM appears to recognize when music uses the classical genre.
This appearance evaporates when we see in Table \ref{tab:SRCAMClassicalConfusions}
what Classical excerpts SRCAM confuses for Blues,
Country, Disco, Jazz, Metal, and Rock,
and which non-Classical excerpts SRCAM confuses for Classical.
This argues the contrary that SRCAM has some capacity to recognize 
when music uses the classical genre.
(We see the same kinds of catastrophic failures for Metal,
the other class for which SRCAM appears to do well.)
%As an example, SRCAM confuses labels as Classical the 
%1984 track of British synthpop group Bonski Beat;
%and it classifies different movements of 
%Henri Dutilleux's String Quartet as Blues and Country.
%The only class SRCAM does not use is Hip hop.

\begin{table*}[t]
\centering
\scriptsize
\begin{tabular}{|@{\;}c@{\;}|@{\;}p{2in}@{\;}|@{\;}p{2.4in}@{\;}|}
\hline 
& {\bf Non-Classical classified as Classical} & {\bf Classical misclassifications} \\ \hline 
Blues     & ``Sugar Mama'' John Lee Hooker (5) & ``Ainsi la nuit for String Quartet: VII'' Henri Dutilleux (41); ``Fuge F\"ur Klavier'' Richard Strauss (45) \\\hline
Country & ``San Antonio Rose'' Floyd Cramer (17); ``My Heroes Have Always Been Cowboys'' Willie Nelson (69) & ``Ainsi la nuit for String Quartet: I'' Henri Dutilleux (38) \\\hline
Disco & ``WHY?'' Bronski Beat (94) & ``Konzert F\"ur Waldhorn Mit Orchester, Allegro'' Richard Strauss (43); ``Violin Concerto No. 1'' Karol Szymanowski (46) \\\hline
Jazz & ``'Round Midnight'' James Carter (2); ``You Never Told Me That You Care'' James Carter (3);
``My Blue Heaven'' Coleman Hawkins (41); ``There Will Never Be Another You'' Coleman Hawkins (57) & 
unidentified (61); ``Piano Sonata No.21 in B flat, D.960, Scherzo'' Franz Schubert (65) \\\hline
Metal & & ``Solemn Mass for Feast of Santa Maria della Salute'' Giovanni Rovetta (56) \\\hline
Pop & ``I want you to need me'' Celine Dion (40) & \\\hline
Reggae & ``Reggae Hit the Town'' The Ethiopians (43) & \\\hline
Rock & ``The Song Remains The Same'' Led Zeppelin (39) & 
``Symphony 6, mvt. 1'' Tchaikovsky (51); ``Candide Overture'' Leonard Bernstein (52) \\ \hline
\end{tabular}
\caption{For SRCAM, confusions as Classical, and confusions of Classical. Excerpt numbers in parentheses.}
\label{tab:SRCAMClassicalConfusions}
%\vspace{-0.35in}
\end{table*}

\section{Conclusions on the future use of {\em GTZAN}}
%\vspace{-0.05in}
It should now be incontrovertible that since all 96 systems in Fig. \ref{fig:GTZANaccs} are evaluated 
using {\em GTZAN} in ST,
we are unable to judge which is good at reproducing the labels in {\em GTZAN}
when efforts are taken to deal with the faults of {\em GTZAN},
let alone which, if any, has any capacity to recognize genres used by music in the real world,
{\em and is thus useful for MGR}.
MAPsCAT and SRCAM,
previously evaluated in {\em GTZAN} to 
have classification accuracies of 
83\% using ST \cite{Sturm2012c,Sturm2012e},
%--- 
%%above the median accuracy of 78.9\%, and 
%superior to 56 others in Fig. \ref{fig:GTZANaccs} ---
now sit at the bottom of Fig. \ref{fig:GTZANaccs}.
Where all the others lie, we do not yet know;
but we now know that these were two systems
with classification accuracies superior to 56 others in Fig. \ref{fig:GTZANaccs}.
Furthermore, the picture becomes more bleak when we scratch below the surface:
for the class in which its figures of merit are the best,
SRCAM shows behaviors that are utterly confusing 
if it really is recognizing that genre.

One might now hold little hope that
{\em GTZAN} has ever been or could ever be useful for
tasks such as evaluating systems for MGR, audio similarity,
autotagging, etc.
However, we have done just that in our analysis above,
as well as in previous work \cite{Sturm2012c,Sturm2012e}.
A proper analysis of the results of an MGR system tested on {\em GTZAN}
must take care of the content of {\em GTZAN}, i.e., the music.
%To determine whether a system has a capacity to recognize genre,
%classification accuracy is not enough, and
%recall, precision, and the F-score are still not enough \cite{Sturm2012e}.
It is not a question of what figure of merit to use,
but of how to draw a valid conclusion with an experimental design
that determines whether the decisions and behaviors of a system
are related to the musical content {\em that is supposedly behind those decisions} \cite{Sturm2013e}.

Some argue that our litany of faults 
above ignores what they say is the
most serious problem with {\em GTZAN}:
that it is too small to produce meaningful results.
In some respects, this is justified.
While personal music collections may number 
thousands of pieces of music, commercial datasets
and library archives number in the millions.
The 1000 excerpts of {\em GTZAN} can certainly be argued
an insufficient random sample of the population of excerpts
``exemplary'' of the genres between which one wishes
a useful MGR system to discriminate.
Hence, one might argue, it is unreasonably optimistic
to assume an MGR system can learn from a fraction of {\em GTZAN} 
those characteristics common to particular genres.
It is beyond the scope of this paper
whether or not a system can learn from {\em GTZAN}
the meaning of the descriptors we see in Fig. \ref{fig:GTZANtags}.
However, though a dataset may be larger and
more modern than {\em GTZAN}, does not mean
it is free of the same kinds of faults we find in {\em GTZAN}.
At least with {\em GTZAN},
one now has a manageable, public, and finally well-studied dataset, 
which is now new and improved with metadata.

As one final comment, that a dataset is large does not free
the creator of an MGR system of the necessarily difficult task of designing, implementing,
and analyzing an evaluation having the validity to conclude
how well the system solves or even addresses
the problem of MGR. %\footnote{http://www.youtube.com/watch?v=a1Y73sPHKxw}
In fact, no valid and meaningful conclusion can come from an evaluation of an MGR system
using {\em Classify} in {\em GTZAN}, or for that matter, 
any dataset having uncontrolled independent variables \cite{Sturm2012e,Sturm2013e}.
Other approaches to evaluation are necessary,
and luckily, there are many alternatives \cite{Sturm2012d}.

\scriptsize
\section*{\small Acknowledgments}
Thanks to: Mark Levy for helpful discussions about {\tt last.fm} tags;
Mads G. Christensen, Nick Collins, Cynthia Liem,
and Clemens Hage for helping identify several excerpts in {\em GTZAN};
Fabien Gouyon for illuminating discussions about these topics;
Carla Sturm for bearing my repeated playing of all excerpts;
and to the many anonymous reviewers for many comments
that helped move this paper nearer to publishability.

\bibliographystyle{IEEEbib}

\begin{thebibliography}{100}

\bibitem{Sturm2012d}
B.~L. Sturm,
\newblock ``A survey of evaluation in music genre recognition,''
\newblock in {\em Proc. Adaptive Multimedia Retrieval}, Copenhagen, Denmark,
  Oct. 2012.

\bibitem{Ahonen2010}
T.~E. Ahonen,
\newblock ``Compressing lists for audio classification,''
\newblock in {\em Proc. Int. Workshop Machine Learning and Music}, 2010, pp.
  45--48.

\bibitem{Anden2011}
J.~And{\'e}n and S.~Mallat,
\newblock ``Multiscale scattering for audio classification,''
\newblock in {\em Proc. ISMIR}, 2011, pp. 657--662.

\bibitem{Anglade2010}
A.~Anglade, E.~Benetos, M.~Mauch, and S.~Dixon,
\newblock ``Improving music genre classification using automatically induced
  harmony rules,''
\newblock {\em J. New Music Research}, vol. 39, no. 4, pp. 349--361, 2010.

\bibitem{Arabi2009}
A.~F. Arabi and G.~Lu,
\newblock ``Enhanced polyphonic music genre classification using high level
  features,''
\newblock in {\em IEEE Int. Conf. Signal and Image Process. App.}, 2009.

\bibitem{Ariyaratne2012}
H.~B. Ariyaratne and D.~Zhang,
\newblock ``A novel automatic hierachical approach to music genre
  classification,''
\newblock in {\em Proc. ICME}, July 2012, pp. 564 --569.

\bibitem{Bagci2007}
U.~Ba\u{g}ci and E.~Erzin,
\newblock ``Automatic classification of musical genres using inter-genre
  similarity,''
\newblock {\em IEEE Signal Proc. Letters}, vol. 14, no. 8, pp. 521--524, Aug.
  2007.

\bibitem{Barreira2011}
L.~Barreira, S.~Cavaco, and J.~da~Silva,
\newblock ``Unsupervised music genre classification with a model-based
  approach,''
\newblock in {\em Proc. Portugese Conf. Progress Artificial Intell.}, 2011, pp.
  268--281.

\bibitem{Behun2012}
K.~Behun,
\newblock ``Image features in music style recognition,''
\newblock in {\em Proc. Central European Seminar on Computer Graphics}, 2012.

\bibitem{Benetos2008}
E.~Benetos and C.~Kotropoulos,
\newblock ``A tensor-based approach for automatic music genre classification,''
\newblock in {\em Proc. EUSIPCO}, Lausanne, Switzerland, 2008.

\bibitem{Benetos2010}
E.~Benetos and C.~Kotropoulos,
\newblock ``Non-negative tensor factorization applied to music genre
  classification,''
\newblock {\em IEEE Trans. Audio, Speech, Lang. Process.}, vol. 18, no. 8, pp.
  1955--1967, Nov. 2010.

\bibitem{Bergstra2006}
J.~Bergstra, N.~Casagrande, D.~Erhan, D.~Eck, and B.~K\'{e}gl,
\newblock ``Aggregate features and {AdaBoost} for music classification,''
\newblock {\em Machine Learning}, vol. 65, no. 2-3, pp. 473--484, June 2006.

\bibitem{Bergstra2006b}
J.~Bergstra,
\newblock ``Algorithms for classifying recorded music by genre,''
\newblock M.S. thesis, Universit\'e de Montr\'eal, Montr\'eal, Canada, Aug.
  2006.

\bibitem{Bergstra2010}
J.~Bergstra, M.~Mandel, and D.~Eck,
\newblock ``Scalable genre and tag prediction with spectral covariance,''
\newblock in {\em Proc. ISMIR}, 2010.

\bibitem{Bogdanov2011}
D.~Bogdanov, J.~Serra, N.~Wack, P.~Herrera, and X.~Serra,
\newblock ``Unifying low-level and high-level music similarity measures,''
\newblock {\em IEEE Trans. Multimedia}, vol. 13, no. 4, pp. 687--701, Aug.
  2011.

\bibitem{Chang2010}
K.~Chang, J.-S.~R. Jang, and C.~S. Iliopoulos,
\newblock ``Music genre classification via compressive sampling,''
\newblock in {\em Proc. ISMIR}, Amsterdam, The Netherlands, Aug. 2010, pp.
  387--392.

\bibitem{Chathuranga2012}
D.~Chathuranga and L.~Jayaratne,
\newblock ``Musical genre classification using ensemble of classifiers,''
\newblock in {\em Proc. Int. Conf. Computational Intelligence, Modelling and
  Simulation}, Sep. 2012, pp. 237--242.

\bibitem{Chen2006}
K.~Chen, S.~Gao, Y.~Zhu, and Q.~Sun,
\newblock ``Music genres classification using text categorization method,''
\newblock in {\em Proc. IEEE Workshop Multimedia Signal Process.}, Oct. 2006,
  pp. 221--224.

\bibitem{Chen2008}
G.~Chen, T.~Wang, and P.~Herrera,
\newblock ``Relevance feedback in an adaptive space with one-class {SVM} for
  content-based music retrieval,''
\newblock in {\em Proc. ICALIP}, July 2008, pp. 1153--1158.

\bibitem{Dixon2010}
S.~Dixon, M.~Mauch, and A.~Anglade,
\newblock ``Probabilistic and logic-based modelling of harmony,''
\newblock in {\em Proc. CMMR}, 2010.

\bibitem{Draman2011}
N.~A. Draman, S.~Ahmad, and A.~K. Muda,
\newblock ``Recognizing patterns of music signals to songs classification using
  modified {AIS}-based classifier,''
\newblock in {\em Software Engineering and Computer Systems}, pp. 724--737.
  Springer Berlin / Heidelberg, 2011.

\bibitem{Fu2010}
Z.~Fu, G.~Lu, K.~M. Ting, and D.~Zhang,
\newblock ``Learning naive {B}ayes classifiers for music classification and
  retrieval,''
\newblock in {\em Proc. ICPR}, 2010, pp. 4589--4592.

\bibitem{Fu2010b}
Z.~Fu, G.~Lu, K.~M. Ting, and D.~Zhang,
\newblock ``On feature combination for music classification,''
\newblock in {\em Proc. Int. Workshop Structural and Syntactic Patt. Recog.},
  2010, pp. 453--462.

\bibitem{Fu2011b}
Z.~Fu, G.~Lu, K.~M. Ting, and D.~Zhang,
\newblock ``Music classification via the bag-of-features approach,''
\newblock {\em Patt. Recgn. Lett.}, vol. 32, no. 14, pp. 1768--1777, Oct. 2011.

\bibitem{Genussov2010}
M.~Genussov and I.~Cohen,
\newblock ``Musical genre classification of audio signals using geometric
  methods,''
\newblock in {\em Proc. EUSIPCO}, Aalborg, Denmark, Aug. 2010, pp. 497--501.

\bibitem{Guaus2009}
E.~Guaus,
\newblock {\em Audio content processing for automatic music genre
  classification: descriptors, databases, and classifiers},
\newblock Ph.D. thesis, Universitat Pompeu Fabra, Barcelona, Spain, 2009.

\bibitem{Hamel2010}
P.~Hamel and D.~Eck,
\newblock ``Learning features from music audio with deep belief networks,''
\newblock in {\em Proc. ISMIR}, 2010.

\bibitem{Hartmann2011}
M.~A. Hartmann,
\newblock ``Testing a spectral-based feature set for audio genre
  classification,''
\newblock M.S. thesis, University of Jyv\"askyl\"a, June 2011.

\bibitem{Henaff2011}
M.~Henaff, K.~Jarrett, K.~Kavukcuoglu, and Y.~LeCun,
\newblock ``Unsupervised learning of sparse features for scalable audio
  classification,''
\newblock in {\em Proc. ISMIR}, Miami, FL, Oct. 2011.

\bibitem{Holzapfel2007}
A.~Holzapfel and Y.~Stylianou,
\newblock ``A statistical approach to musical genre classification using
  non-negative matrix factorization,''
\newblock in {\em Proc. ICASSP}, Apr. 2007, pp. 693--696.

\bibitem{Holzapfel2008}
A.~Holzapfel and Y.~Stylianou,
\newblock ``Musical genre classification using nonnegative matrix
  factorization-based features,''
\newblock {\em IEEE Trans. Audio, Speech and Lang. Process.}, vol. 16, no. 2,
  pp. 424--434, Feb. 2008.

\bibitem{Karkavitsas2011}
George~V. Karkavitsas and George~A. Tsihrintzis,
\newblock ``Automatic music genre classification using hybrid genetic
  algorithms,''
\newblock in {\em Intelligent Interactive Multimedia Systems and Services}, pp.
  323--335. Springer Berlin / Heidelberg, 2011.

\bibitem{Karkavitsas2012}
G.~V. Karkavitsas and F.~A. Tsihrintzis,
\newblock ``Optimization of an automatic music genre classification system via
  hyper-entities,''
\newblock in {\em Proc. Int. Conf. Intell. Info. Hiding and Multimedia Signal
  Process.}, 2012, pp. 449--452.

\bibitem{Kotropoulos2010}
C.~Kotropoulos, G.~R. Arce, and Y.~Panagakis,
\newblock ``Ensemble discriminant sparse projections applied to music genre
  classification,''
\newblock in {\em Proc. ICPR}, Aug. 2010, pp. 823--825.

\bibitem{Krasser2012}
J.~Krasser, J.~Abe{\ss}er, H.~Gro{\ss}mann, C.~Dittmar, and E.~Cano,
\newblock ``Improved music similarity computation based on tone objects,''
\newblock in {\em Proc. Audio Mostly Conf.}, 2012, pp. 47--54.

\bibitem{Lampropoulos2010}
A.~S. Lampropoulos, P.~S. Lampropoulou, and G.~A. Tsihrintzis,
\newblock ``Music genre classification based on ensemble of signals produced by
  source separation methods,''
\newblock {\em Intelligent Decision Technologies}, vol. 4, no. 3, pp. 229--237,
  2010.

\bibitem{Lee2009b}
C.~Lee, J.~Shih, K.~Yu, and H.~Lin,
\newblock ``Automatic music genre classification based on modulation spectral
  analysis of spectral and cepstral features,''
\newblock {\em IEEE Trans. Multimedia}, vol. 11, no. 4, pp. 670--682, June
  2009.

\bibitem{Leon2012}
F.~de~Leon and K.~Martinez,
\newblock ``Enhancing timbre model using mfcc and its time derivatives for
  music similarity estimation,''
\newblock in {\em Proc. EUSIPCO}, Bucharest, Romania, Aug. 2012, pp.
  2005--2009.

\bibitem{Leon2012b}
F.~de~Leon and K.~Martinez,
\newblock ``Towards efficient music genre classification using {FastMap},''
\newblock in {\em Proc. DAFx}, 2012.

\bibitem{Li2003}
T.~Li, M.~Ogihara, and Q.~Li,
\newblock ``A comparative study on content-based music genre classification,''
\newblock in {\em Proc. ACM SIGIR}, 2003.

\bibitem{Li2003b}
T.~Li and G.~Tzanetakis,
\newblock ``Factors in automatic musical genre classification of audio
  signals,''
\newblock in {\em Proc. IEEE Workshop Appl. Signal Process. Audio Acoust.},
  2003.

\bibitem{Li2005}
M.~Li and R.~Sleep,
\newblock ``Genre classification via an {LZ78}-based string kernel,''
\newblock in {\em Proc. ISMIR}, 2005.

\bibitem{Li2005b}
T.~Li and M.~Ogihara,
\newblock ``Music genre classification with taxonomy,''
\newblock in {\em Proc. ICASSP}, Philadelphia, PA, Mar. 2005, pp. 197--200.

\bibitem{Li2006}
T.~Li and M.~Ogihara,
\newblock ``Toward intelligent music information retrieval,''
\newblock {\em IEEE Trans. Multimedia}, vol. 8, no. 3, pp. 564--574, June 2006.

\bibitem{Li2010}
T.~LH. Li, A.~B. Chan, and A.~HW. Chun,
\newblock ``Automatic musical pattern feature extraction using convolutional
  neural network,''
\newblock in {\em Proc. Int. Conf. Data Mining and Applications}, 2010.

\bibitem{Li2011}
T.~Li and A.~Chan,
\newblock ``Genre classification and the invariance of {MFCC} features to key
  and tempo,''
\newblock in {\em Proc. Int. Conf. MultiMedia Modeling}, Taipei, China, Jan.
  2011.

\bibitem{Lidy2005}
T.~Lidy and A.~Rauber,
\newblock ``Evaluation of feature extractors and psycho-acoustic
  transformations for music genre classification,''
\newblock in {\em Proc. ISMIR}, 2005.

\bibitem{Lidy2006}
T.~Lidy,
\newblock ``Evaluation of new audio features and their utilization in novel
  music retrieval applications,''
\newblock M.S. thesis, Vienna University of Tech., December 2006.

\bibitem{Lidy2007}
T.~Lidy, A.~Rauber, A.~Pertusa, and J.~M.~I\ nesta,
\newblock ``Improving genre classification by combination of audio and symbolic
  descriptors using a transcription system,''
\newblock in {\em Proc. ISMIR}, Vienna, Austria, Sep. 2007, pp. 61--66.

\bibitem{Lidy2010b}
T.~Lidy, R.~Mayer, A.~Rauber, P.~P. de~Leon, A.~Pertusa, and J.~Quereda,
\newblock ``A cartesian ensemble of feature subspace classifiers for music
  categorization,''
\newblock in {\em Proc. ISMIR}, 2010, pp. 279--284.

\bibitem{Lim2011}
S.-C. Lim, S.-J. Jang, S.-P. Lee, and M.~Y. Kim,
\newblock ``Music genre/mood classification using a feature-based modulation
  spectrum,''
\newblock in {\em Proc. Int. Conf. Modelling, Identification and Control},
  2011.

\bibitem{Liu2009}
Y.~Liu, L.~Wei, and P.~Wang,
\newblock ``Regional style automatic identification for {Chinese} folk songs,''
\newblock in {\em World Cong. Computer Science and Information Engineering},
  2009.

\bibitem{Manzagol2008}
P.-A Manzagol, T.~Bertin-Mahieux, and D.~Eck,
\newblock ``On the use of sparse time-relative auditory codes for music,''
\newblock in {\em Proc. ISMIR}, Philadelphia, PA, Sep. 2008, pp. 603--608.

\bibitem{Markov2012}
K.~Markov and T.~Matsui,
\newblock ``Music genre classification using self-taught learning via sparse
  coding,''
\newblock in {\em Proc. ICASSP}, Mar. 2012, pp. 1929 --1932.

\bibitem{Markov2012b}
K.~Markov and T.~Matsui,
\newblock ``Nonnegative matrix factorization based self-taught learning with
  application to music genre classification,''
\newblock in {\em Proc. IEEE Int. Workshop Machine Learn. Signal Process.},
  Sep. 2012, pp. 1--5.

\bibitem{Marques2011c}
C.~Marques, I.~R. Guiherme, R.~Y.~M. Nakamura, and J.~P. Papa,
\newblock ``New trends in musical genre classification using optimum-path
  forest,''
\newblock in {\em Proc. ISMIR}, 2011.

\bibitem{Mayer2010c}
R.~Mayer, A.~Rauber, P.~J. Ponce~de Le\'{o}n, C.~P{\'e}rez-Sancho, and J.~M.
  I\~{n}esta,
\newblock ``Feature selection in a cartesian ensemble of feature subspace
  classifiers for music categorisation,''
\newblock in {\em Proc. ACM Int. Workshop Machine Learning and Music}, 2010,
  pp. 53--56.

\bibitem{Moerchen2006}
F.~Moerchen, I.~Mierswa, and A.~Ultsch,
\newblock ``Understandable models of music collections based on exhaustive
  feature generation with temporal statistics,''
\newblock in {\em Int. Conf. Knowledge Discover and Data Mining}, 2006.

\bibitem{Nagathil2011}
A.~Nagathil, P.~G\"ottel, and R.~Martin,
\newblock ``Hierarchical audio classification using cepstral modulation ratio
  regressions based on legendre polynomials,''
\newblock in {\em Proc. ICASSP}, July 2011, pp. 2216--2219.

\bibitem{Panagakis2008}
Y.~Panagakis, E.~Benetos, and C.~Kotropoulos,
\newblock ``Music genre classification: A multilinear approach,''
\newblock in {\em Proc. ISMIR}, Philadelphia, PA, Sep. 2008, pp. 583--588.

\bibitem{Panagakis2009}
Y.~Panagakis, C.~Kotropoulos, and G.~R. Arce,
\newblock ``Music genre classification via sparse representations of auditory
  temporal modulations,''
\newblock in {\em Proc. EUSIPCO}, Aug. 2009.

\bibitem{Panagakis2009b}
Y.~Panagakis, C.~Kotropoulos, and G.~R. Arce,
\newblock ``Music genre classification using locality preserving non-negative
  tensor factorization and sparse representations,''
\newblock in {\em Proc. ISMIR}, Kobe, Japan, Oct. 2009, pp. 249--254.

\bibitem{Panagakis2010}
Y.~Panagakis, C.~Kotropoulos, and G.~R. Arce,
\newblock ``Non-negative multilinear principal component analysis of auditory
  temporal modulations for music genre classification,''
\newblock {\em IEEE Trans. Audio, Speech, Lang. Process.}, vol. 18, no. 3, pp.
  576--588, Mar. 2010.

\bibitem{Panagakis2010c}
Y.~Panagakis and C.~Kotropoulos,
\newblock ``Music genre classification via topology preserving non-negative
  tensor factorization and sparse representations,''
\newblock in {\em Proc. ICASSP}, Mar. 2010, pp. 249--252.

\bibitem{Ravelli2010}
E.~Ravelli, G.~Richard, and L.~Daudet,
\newblock ``Audio signal representations for indexing in the transform
  domain,''
\newblock {\em IEEE Trans. Audio, Speech, Lang. Process.}, vol. 18, no. 3, pp.
  434--446, Mar. 2010.

\bibitem{Ren2011}
J.-M. Ren and J.-S.~R. Jang,
\newblock ``Time-constrained sequential pattern discovery for music genre
  classification,''
\newblock in {\em Proc. ICASSP}, 2011, pp. 173--176.

\bibitem{Ren2012}
J.-M. Ren and J.-S.~R. Jang,
\newblock ``Discovering time-constrained sequential patterns for music genre
  classification,''
\newblock {\em IEEE Trans. Audio, Speech, and Lang. Process.}, vol. 20, no. 4,
  pp. 1134--1144, May 2012.

\bibitem{Rocha2011}
B.~Rocha,
\newblock ``Genre classification based on predominant melodic pitch contours,''
\newblock M.S. thesis, Universitat Pompeu Fabra, Barcelona, Spain, Sep. 2011.

\bibitem{Rump2010}
H.~Rump, S.~Miyabe, E.~Tsunoo, N.~Ono, and S.~Sagayama,
\newblock ``Autoregressive {MFCC} models for genre classification improved by
  harmonic-percussion separation,''
\newblock in {\em Proc. ISMIR}, 2010, pp. 87--92.

\bibitem{Salamon2012}
J.~Salamon, B.~Rocha, and E.~Gomez,
\newblock ``Musical genre classification using melody features extracted from
  polyphonic music signals,''
\newblock in {\em Proc. ICASSP}, Kyoto, Japan, Mar. 2012.

\bibitem{Santos2010}
C.~A. de~los Santos,
\newblock ``Nonlinear audio recurrence analysis with application to music genre
  classification,''
\newblock M.S. thesis, Universitat Pompeu Fabra, Barcelona, Spain, 2010.

\bibitem{Schindler2012b}
A.~Schindler and A.~Rauber,
\newblock ``Capturing the temporal domain in echonest features for improved
  classification effectiveness,''
\newblock in {\em Proc. Adaptive Multimedia Retrieval}, Oct. 2012.

\bibitem{Seo2011}
J.~S. Seo and S.~Lee,
\newblock ``Higher-order moments for musical genre classification,''
\newblock {\em Signal Process.}, vol. 91, no. 8, pp. 2154--2157, 2011.

\bibitem{Serra2011}
J.~Serra, C.~A. de~los Santos, and R.~G. Andrzejak,
\newblock ``Nonlinear audio recurrence analysis with application to genre
  classification,''
\newblock in {\em Proc. ICASSP}, 2011.

\bibitem{Seyerlehner2010}
K.~Seyerlehner,
\newblock {\em Content-based Music Recommender Systems: Beyond Simple
  Frame-level Audio Similarity},
\newblock Ph.D. thesis, Johannes Kepler University, Linz, Austria, Dec. 2010.

\bibitem{Seyerlehner2010b}
K.~Seyerlehner, G.~Widmer, and T.~Pohle,
\newblock ``Fusing block-level features for music similarity estimation,''
\newblock in {\em DAFx}, 2010.

\bibitem{Seyerlehner2012}
K.~Seyerlehner, M.~Schedl, R.~Sonnleitner, D.~Hauger, and B.~Ionescu,
\newblock ``From improved auto-taggers to improved music similarity measures,''
\newblock in {\em Proc. Adaptive Multimedia Retrieval}, Copenhagen, Denmark,
  Oct. 2012.

\bibitem{Shen2005}
J.~Shen, J.~Shepherd, and A.~Ngu,
\newblock ``On efficient music genre classification,''
\newblock in {\em Database Systems for Advanced Applications}, pp. 990--990.
  Springer Berlin / Heidelberg, 2005.

\bibitem{Shen2006}
J.~Shen, J.~Shepherd, and A.~H.~H. Ngu,
\newblock ``Towards effective content-based music retrieval with multiple
  acoustic feature combination,''
\newblock {\em IEEE Trans. Multimedia}, vol. 8, no. 6, pp. 1179--1189, Dec.
  2006.

\bibitem{Sotiropoulos2008}
D.~Sotiropoulos, A.~Lampropoulos, and G.~Tsihrintzis,
\newblock ``Artificial immune system-based music genre classification,''
\newblock in {\em New Directions in Intelligent Interactive Multimedia}, pp.
  191--200. Springer Berlin / Heidelberg, 2008.

\bibitem{Srinivasan2004}
H.~Srinivasan and M.~Kankanhalli,
\newblock ``Harmonicity and dynamics-based features for audio,''
\newblock in {\em Proc. ICASSP}, May 2004, vol.~4, pp. 321--324.

\bibitem{Sturm2012}
B.~L. Sturm and P.~Noorzad,
\newblock ``On automatic music genre recognition by sparse representation
  classification using auditory temporal modulations,''
\newblock in {\em Proc. CMMR}, London, UK, June 2012.

\bibitem{Sturm2012c}
B.~L. Sturm,
\newblock ``Two systems for automatic music genre recognition: What are they
  really recognizing?,''
\newblock in {\em Proc. ACM MIRUM Workshop}, Nara, Japan, Nov. 2012.

\bibitem{Sturm2012e}
B.~L. Sturm,
\newblock ``Classification accuracy is not enough: On the evaluation of music
  genre recognition systems,''
\newblock {\em J. Intell. Info. Systems (accepted)}, 2013.

\bibitem{Sturm2013}
B.~L. Sturm,
\newblock ``On music genre classification via compressive sampling,''
\newblock in {\em Proc. ICME}, 2013.

\bibitem{Sturm2013b}
B.~L. Sturm,
\newblock ``Music genre recognition with risk and rejection,''
\newblock in {\em Proc. ICME}, 2013.

\bibitem{Tietche2012}
B.~H. Tietche, O.~Romain, B.~Denby, L.~Benaroya, and S.~Viateur,
\newblock ``{FPGA}-based radio-on-demand broadcast receiver with musical genre
  identification,''
\newblock in {\em Proc. IEEE Int. Symp. Industrial Elect.}, May 2012, pp.
  1381--1385.

\bibitem{Tsunoo2009}
E.~Tsunoo, G.~Tzanetakis, N.~Ono, and S.~Sagayama,
\newblock ``Audio genre classification by clustering percussive patterns,''
\newblock in {\em Proc. Acoustical Society of Japan}, 2009.

\bibitem{Tsunoo2009b}
E.~Tsunoo, G.~Tzanetakis, N.~Ono, and S.~Sagayama,
\newblock ``Audio genre classification using percussive pattern clustering
  combined with timbral features,''
\newblock in {\em Proc. ICME}, 2009.

\bibitem{Tsunoo2011}
E.~Tsunoo, G.~Tzanetakis, N.~Ono, and S.~Sagayama,
\newblock ``Beyond timbral statistics: Improving music classification using
  percussive patterns and bass lines,''
\newblock {\em IEEE Trans. Audio, Speech, and Lang. Process.}, vol. 19, no. 4,
  pp. 1003--1014, May 2011.

\bibitem{Turnbull2005}
D.~Turnbull and C.~Elkan,
\newblock ``Fast recognition of musical genres using {RBF} networks,''
\newblock {\em IEEE Trans. Knowl. Data Eng.}, vol. 17, no. 4, pp. 580--584,
  Apr. 2005.

\bibitem{Tzanetakis2002}
G.~Tzanetakis and P.~Cook,
\newblock ``Musical genre classification of audio signals,''
\newblock {\em IEEE Trans. Speech Audio Process.}, vol. 10, no. 5, pp.
  293--302, July 2002.

\bibitem{Tzanetakis2002b}
G.~Tzanetakis,
\newblock {\em Manipulation, Analysis and Retrieval Systems for Audio Signals},
\newblock Ph.D. thesis, Princeton University, June 2002.

\bibitem{Wu2011}
M.-J. Wu, Z.-S. Chen, J.-S.~R. Jang, and J.-M. Ren,
\newblock ``Combining visual and acoustic features for music genre
  classification,''
\newblock in {\em Int. Conf. Machine Learning and Applications}, 2011.

\bibitem{Wulfing2012}
J.~W\"ulfing and M.~Riedmiller,
\newblock ``Unsupervised learning of local features for music classification,''
\newblock in {\em Proc. ISMIR}, Porto, Portugal, Oct. 2012.

\bibitem{Yang2011b}
X.~Yang, Q.~Chen, S.~Zhou, and X.~Wang,
\newblock ``Deep belief networks for automatic music genre classification,''
\newblock in {\em Proc. INTERSPEECH}, 2011, pp. 2433--2436.

\bibitem{Yaslan2006b}
Y.~Yaslan and Z.~Cataltepe,
\newblock ``Audio music genre classification using different classifiers and
  feature selection methods,''
\newblock in {\em Proc. ICPR}, 2006, pp. 573--576.

\bibitem{Yeh2012}
C.-C.~M. Yeh and Y.-H. Yang,
\newblock ``Supervised dictionary learning for music genre classification,''
\newblock in {\em Proc. ACM Int. Conf. Multimedia Retrieval}, Hong Kong, China,
  Jun. 2012.

\bibitem{Yeh2013}
C.-C.~M. Yeh, L.~Su, and Y.-H. Yang,
\newblock ``Dual-layer bag-of-frames model for music genre classification,''
\newblock in {\em Proc. ICASSP}, 2013.

\bibitem{Zeng2009}
Z.~Zeng, S.~Zhang, H.~Li, W.~Liang, and H.~Zheng,
\newblock ``A novel approach to musical genre classification using
  probabilistic latent semantic analysis model,''
\newblock in {\em Proc. ICME}, 2009, pp. 486--489.

\bibitem{Zhou2012}
G.-T. Zhou, K.~M. Ting, F.~T. Liu, and Y.~Yin,
\newblock ``Relevance feature mapping for content-based multimedia information
  retrieval,''
\newblock {\em Patt. Recog.}, vol. 45, pp. 1707--1720, 2012.

\bibitem{Fu2011}
Z.~Fu, G.~Lu, K.~M. Ting, and D.~Zhang,
\newblock ``A survey of audio-based music classification and annotation,''
\newblock {\em IEEE Trans. Multimedia}, vol. 13, no. 2, pp. 303--319, Apr.
  2011.

\bibitem{Sturm2012b}
B.~L. Sturm,
\newblock ``An analysis of the {GTZAN} music genre dataset,''
\newblock in {\em Proc. ACM MIRUM Workshop}, Nara, Japan, Nov. 2012.

\bibitem{Urbano2011b}
J.~Urbano,
\newblock ``Information retrieval meta-evaluation: Challenges and opportunities
  in the music domain,''
\newblock in {\em ISMIR}, 2011, pp. 609--614.

\bibitem{Sturm2013e}
B.~L. Sturm,
\newblock ``Evaluating music emotion recognition: Lessons from music genre
  recognition?,''
\newblock in {\em Proc. ICME}, 2013.

\bibitem{McKay2006}
C.~McKay and I.~Fujinaga,
\newblock ``Music genre classification: Is it worth pursuing and how can it be
  improved?,''
\newblock in {\em Proc. ISMIR}, Victoria, Canada, Oct. 2006.

\bibitem{Craft2007}
A.~Craft, G.~A. Wiggins, and T.~Crawford,
\newblock ``How many beans make five? {T}he consensus problem in music-genre
  classification and a new evaluation method for single-genre categorisation
  systems,''
\newblock in {\em Proc. ISMIR}, 2007.

\bibitem{Craft2007b}
A.~Craft,
\newblock ``The role of culture in the music genre classification task: human
  behaviour and its effect on methodology and evaluation,''
\newblock Tech. {R}ep., Queen Mary University of London, Nov. 2007.

\bibitem{Wiggins2009}
G.~A. Wiggins,
\newblock ``Semantic gap?? {S}chemantic schmap!! {M}ethodological
  considerations in the scientific study of music,''
\newblock in {\em Proc. IEEE Int. Symp. Mulitmedia}, Dec. 2009, pp. 477--482.

\bibitem{Bertin-Mahieux2010c}
T.~Bertin-Mahieux, D.~Eck, and M.~Mandel,
\newblock ``Automatic tagging of audio: The state-of-the-art,''
\newblock in {\em Machine Audition: Principles, Algorithms and Systems},
  W.~Wang, Ed. IGI Publishing, 2010.

\bibitem{Law2011}
E.~Law,
\newblock ``Human computation for music classification,''
\newblock in {\em Music Data Mining}, T.~Li, M.~Ogihara, and G.~Tzanetakis,
  Eds., pp. 281--301. CRC Press, 2011.

\bibitem{Bertin-Mahieux2008}
T.~Bertin-Mahieux, D.~Eck, F.~Maillet, and P.~Lamere,
\newblock ``Autotagger: A model for predicting social tags from acoustic
  features on large music databases,''
\newblock {\em Journal of New Music Research}, vol. 37, no. 2, pp. 115--135,
  2008.

\bibitem{Barrington2008}
Luke Barrington, Mehrdad Yazdani, Douglas Turnbull, and Gert R.~G. Lanckriet,
\newblock ``Combining feature kernels for semantic music retrieval,''
\newblock in {\em ISMIR}, 2008, pp. 614--619.

\bibitem{Ammer2004}
C.~Ammer,
\newblock {\em Dictionary of Music},
\newblock The Facts on File, Inc., New York, NY, USA, 4 edition, 2004.

\bibitem{Shapiro2005}
P.~Shapiro,
\newblock {\em Turn the Beat Around: The Secret History of Disco},
\newblock Faber \& Faber, London, U.K., 2005.

\bibitem{Wang2003}
A.~Wang,
\newblock ``An industrial strength audio search algorithm,''
\newblock in {\em Proc. Int. Soc. Music Info. Retrieval}, Baltimore, Maryland,
  USA, Oct. 2003.

\bibitem{ISMIR2004}
ISMIR,
\newblock ``Genre results,''
  \url{http://ismir2004.ismir.net/genre_contest/index.htm}, 2004.

\bibitem{Sturm2013d}
B.~L. Sturm and F.~Gouyon,
\newblock ``Comments on ``automatic classification of musical genres using
  inter-genre similarity'',''
\newblock 2013 (submitted).

\bibitem{Pampalk2005b}
E.~Pampalk, A.~Flexer, and G.~Widmer,
\newblock ``Improvements of audio-based music similarity and genre
  classification,''
\newblock in {\em Proc. ISMIR}, London, U.K., Sep. 2005, pp. 628--233.

\bibitem{Flexer2007}
A.~Flexer,
\newblock ``A closer look on artist filters for musical genre classification,''
\newblock in {\em Proc. ISMIR}, Vienna, Austria, Sep. 2007.

\bibitem{Flexer2009}
A.~Flexer and D.~Schnitzer,
\newblock ``Album and artist effects for audio similarity at the scale of the
  web,''
\newblock in {\em Proc. SMC}, Porto, Portugal, July 2009, pp. 59--64.

\bibitem{Flexer2010}
A.~Flexer and D.~Schnitzer,
\newblock ``Effects of album and artist filters in audio similarity computed
  for very large music databases,''
\newblock {\em Computer Music J.}, vol. 34, no. 3, pp. 20--28, 2010.

\bibitem{Theodoridis2009}
S.~Theodoridis and K.~Koutroumbas,
\newblock {\em Pattern Recognition},
\newblock Academic Press, Elsevier, Amsterdam, The Netherlands, 4 edition,
  2009.

\bibitem{PRTools}
R.~P.~W. Duin, P.~Juszczak, D.~de~Ridder, P.~Paclik, E.~Pekalska, and D.~M.~J.
  Tax,
\newblock ``{PR}-{T}ools4.1, a matlab toolbox for pattern recognition,'' Delft
  University of Technology, 2007,
\newblock \url{http://prtools.org}.

\bibitem{Slaney1998}
M.~Slaney,
\newblock ``Auditory toolbox,''
\newblock Tech. {R}ep., Interval Research Corporation, 1998.

\bibitem{Wright2009b}
J.~Wright, A.~Y. Yang, A.~Ganesh, S.~S. Sastry, and Y.~Ma,
\newblock ``Robust face recognition via sparse representation,''
\newblock {\em IEEE Trans. Pattern Anal. Machine Intell.}, vol. 31, no. 2, pp.
  210--227, Feb. 2009.

\bibitem{Berg2008}
E.~van~den Berg and M.~P. Friedlander,
\newblock ``Probing the {P}areto frontier for basis pursuit solutions,''
\newblock {\em SIAM J. on Scientific Computing}, vol. 31, no. 2, pp. 890--912,
  Nov. 2008.

\bibitem{Salzberg1997}
S.~L. Salzberg,
\newblock {\em Data mining and knowledge discovery}, chapter On comparing
  classifiers: Pitfalls to avoid and a recommended approach, pp. 317--328,
\newblock Kluwer Academic Publishers, 1997.

\end{thebibliography}

\end{document}